\documentclass[structabstract]{aa}
\usepackage{graphicx}
\usepackage{txfonts}
\usepackage{dcolumn}
\usepackage{natbib}
\begin{document}

\title{Adaptive-optics assisted near-infrared polarization measurements of sources in the Galactic Center}
\titlerunning{AO assisted NIR polarization measurements of sources in the GC}
\author{R. M. Buchholz \inst{1}
       \and 
       G. Witzel \inst{1}
       \and
       R. Sch\"odel \inst{2,1}
       \and 
       A. Eckart \inst{1,3}
       \and
       M. Bremer \inst{1}
       \and
       K. Mu\v{z}i\'{c} \inst{4}
       }

\institute{I. Physikalisches Institut, Universit\"at zu K\"oln,
           Z\"ulpicher Str. 77, 50937 K\"oln, Germany\\
	   \email{buchholz,mbremer,eckart,witzel@ph1.uni-koeln.de}
	   \and Instituto de Astrof\'isica de Andaluc\'ia (CSIC), Glorieta de la Astronom\'ia s/n, E-18008 Granada, Spain\\
	   \email{rainer@iaa.es}
	   \and
           Max-Planck-Institut f\"ur Radioastronomie, 
           Auf dem H\"ugel 69, 53121 Bonn, Germany
	   \and University of Toronto, Department of Astronomy and Astrophysics, 50 St. George Street, Toronto, ON M5S 3H4\\
	   \email{muzic@astro.utoronto.ca}
           }

\date{Received 20 May 2011, accepted 12 July 2011}
\abstract{The Galactic Center offers unique opportunities to study stellar and bow-shock polarization effects in a dusty environment.}
{The goals of this work are to provide near-infrared (NIR) polarimetry of the stellar sources in the central parsec at the
resolution of an 8m telescope for the first time, along with new insights into the nature of the known bright bow-shock sources.}    
{We use adaptive-optics assisted observations obtained at the ESO VLT in the H- and Ks-band, applying both high-precision
photometric methods specifically developed for crowded fields and a newly established polarimetric calibration for NACO to
produce polarization maps of the central 3''$\times$19'', in addition to spatially resolved polarimetry and a flux
variability analysis on the extended sources in this region.}{We find foreground polarization mainly parallel to the
Galactic plane, with average values of (4.6 $\pm$ 0.6)\% at $26^{\circ} \pm 6^{\circ}$ (Ks-band) and (9.3 $\pm$ 1.3)\% at
$20^{\circ} \pm 6^{\circ}$ (H-band) in the center of the field-of-view (FOV). Further away from the center, we find higher polarization
degrees and steeper polarization angles: (7.5 $\pm$ 1.0)\% at $11^{\circ} \pm 6^{\circ}$ (Ks-band) and (12.1 $\pm$ 2.1)\% at
$13^{\circ} \pm 6^{\circ}$ (H-band). $p_H / p_{Ks}$ peaks at $1.9 \pm 0.4$, corresponding to a power law index for the
wavelength dependency of $\alpha = 2.4 \pm 0.7$. These values also vary over the
FOV, with higher values in the center. This is indicative of the influence of local effects on the total polarization,
possibly dichroic extinction by Northern Arm dust. The two extended sources IRS~21 and 1W show similar intrinsic polarization
degrees of 6.1 resp. 7.8\% (Ks) and 6.9 (H, only 1W) at polarization angles coincident with previous NIR and mid-infrared (MIR) findings,
both in total and spatially resolved. The spatial polarization pattern of both sources points to scattering on aligned elongated
dust grains as the major source of intrinsic polarization, and matches the known orientation of the magnetic field. Our data also
allow us to separate the bow shock of IRS~21 from the central source for the first time in the Ks-band, finding the apex north
of the central source and determining a standoff distance of $\sim$400 AU, which matches previous estimates. This source also
shows a $\sim50$\% increase in flux in the NIR over several years.}{}


\maketitle
\section{Introduction}
\begin{figure}[!b]
\centering
\includegraphics[width=\textwidth,angle=0, scale=0.48]{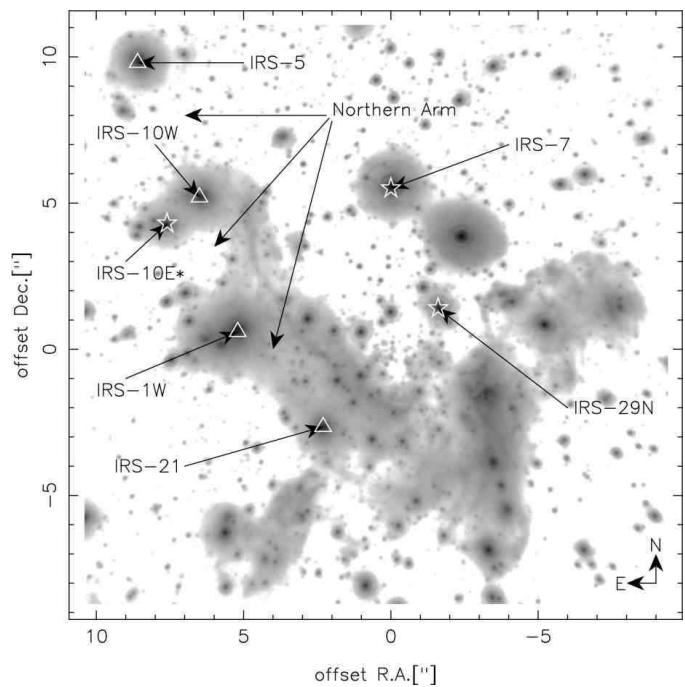}
\caption{\small Lp-band image of the innermost 20'' of the GC (ESO VLT NACO image, program 179.B-0261(A)). Prototypical (bow-shock)
sources and location of the Northern Arm are indicated. Stellar sources are marked with stars, bow-shocks are indicated by triangles.}
\label{FigLband}
\end{figure}

The center of the Milky Way is located at a distance of $\sim8.0$ kpc \citep{ghez2008,gillessen2009}. It contains
a nuclear stellar cluster (NSC) with a $\sim4.0 \times 10^6$ M$_{\sun}$ super-massive black hole at its dynamical
center \citep{eckart2002,schoedel2002,schoedel2003,ghez2003,ghez2008,gillessen2009}. This cluster shows similar
properties as the NSCs found at the dynamical and photometric centers of other galaxies \citep{boeker2010,schoedel2010c}.\\ 
Less than a decade after the first near-infrared (NIR) imaging observations of the Galactic center (GC) by
\cite{becklin_neugebauer1968}, the polarization of 10 sources within $\leq$2 pc of Sagittarius A* was measured by
\cite{capps1976,knacke1977}, observing in the K-, L- and 11.5 $\mu$m-band. These observations revealed similar polarization degrees and
angles for the four sources observed in the K-band, with the polarization angles roughly parallel to the galactic plane
\citep[$\sim$4\% at 15$^{\circ}$ East-of-North, while the projection of the galactic plane is at a position angle of $\sim$31.4$^{\circ}$
at the location of the GC, see ][]{reid2004}. This was interpreted as polarization induced by aligned dust grains in the Milky Way
spiral arms along the line-of-sight (LOS). The 11.5 $\mu$m polarization, however, was found to be almost perpendicular to the
galactic plane and was therefore classified as intrinsic. The values found for the L-band showed intermediate values, which
was attributed to a superposition of both effects. The GC therefore offers the possibility of studying both interstellar
polarization and the properties of intrinsically polarized sources.\\
\cite{kobayashi1980} conducted a survey of K-band polarization in a much wider field of view (7'$\times$7'), finding largely
uniform polarization along the galactic plane. \cite{lebofsky1982} confirmed these findings in the H- and K-band for 17 sources
in the central cluster. The latest H-band survey was conducted by \cite{bailey1984}, who examined $\sim$10
sources with a 4.5'' resolution, finding similar results. While these surveys could only resolve a small number of sources in the
central region, higher resolution observations (0.25'') enabled \cite{eckart1995} to measure the polarization of 160 sources in
the central 13''$\times$13'', while \cite{ott1999} examined $\sim$ 40 bright sources in the central 20''$\times$20'' at 0.5''
resolution. These two studies confirmed the known largely uniform foreground polarization, but already revealed a more complex
picture: individual sources showed different polarization parameters, such as a significantly higher polarization degree (IRS~21).\\
These results already illustrated that in addition to the foreground polarization, intrinsic polarization plays a role in the Ks-band.
Especially sources embedded in the Northern Arm and other bow-shock sources (see Fig.\ref{FigLband} for an overview of the bright
bow-shock sources in the central 20'') show signs of intrinsic polarization. Among these
objects, IRS~21 shows the strongest total Ks-band polarization of a bright GC source detected to date \citep[$\sim$ 10-16\% at 16$^{\circ}$]{eckart1995,ott1999}.
\begin{figure*}[!t]
\centering
\includegraphics[width=\textwidth,angle=0, scale=1.0]{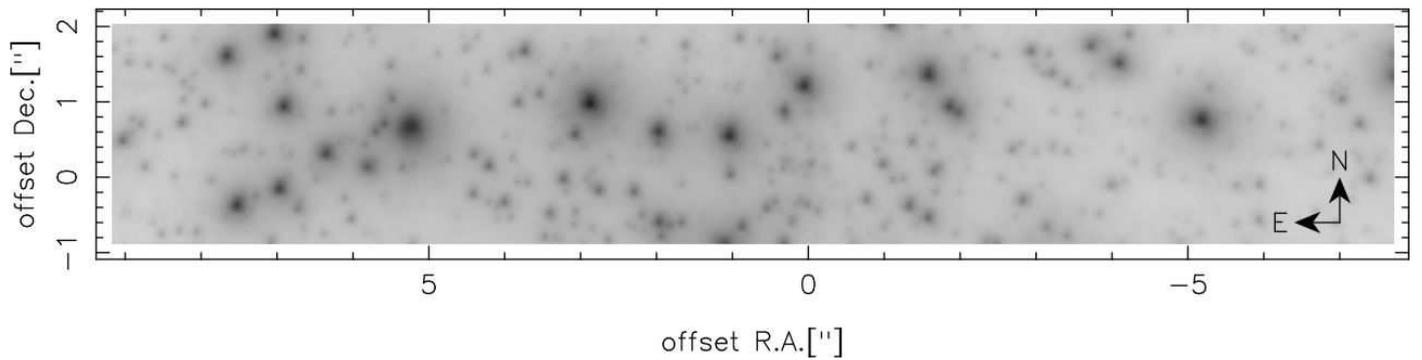}
\caption{\small Ks-band field-of-view, one channel of the Wollaston prism.}
\label{FigFOVwoll}
\end{figure*}
\cite{tanner2002} described IRS~21 as a bow-shock most likely created by a mass-losing Wolf-Rayet star. The observed
polarization is a superposition of foreground polarization and source intrinsic polarization. It is still
unclear what process(es) cause the latter component: (Mie)-scattering in the dusty environment of the Northern
Arm and/or emission from magnetically aligned dust. The latter is known to occur at 12.5 $\mu$m \citep{aitken1998},
but should be negligible at shorter wavelengths.\\
In general, three effects can produce polarized NIR radiation in GC sources: (re)emission by heated, non-spherical
grains, scattering (on spherical and/or aligned non-spherical grains) and dichroic extinction by aligned dust grains.
The first two cases can be regarded as intrinsic to the source for our purposes, thereby allowing conclusions about the
source itself and its immediate environment, while the third effect is the result of grain alignment averaged along the LOS.
For sources enclosed in an optically thick dust shell, dichroic extinction can also contribute significantly as a local
effect \citep[see e.g.][]{whitney2002}.\\
As the basic mechanism that could cause the observed large-scale grain alignment, \cite{davis_greenstein1951} suggested
paramagnetic dissipation, which basically aligns the angular momentum of spinning grains with the magnetic field. But even
almost 60 years later, the problem of grain alignment is by no means completely solved, and it remains difficult to reach exact
conclusions for dust parameters and magnetic field strength, but at least determining the magnetic field orientation is
possible. If the parameters change along the LOS, this further complicates the issue. See e.g. \cite{purcell1971}, \cite{lazarian2003}, cite{lazarian2007}
and references therein for a review of the different possible causes of grain
alignment expected to be relevant in different environments.\\
This makes it possible to use polarimetric measurements to map at least the direction of the magnetic fields responsible for
dust alignment through the Davis-Greenstein effect, as e.g. \cite{nishiyama2009,nishiyama2010} showed for the innermost
20' resp. 2$^{\circ}$ of  the GC, but these studies did not cover the central parsec owing to insufficient resolution.\\
Observations of the Galactic Center suffer from strong extinction caused by dust grains on the LOS, with values of up to
$A_V=40$ mag at optical wavelengths (or even higher values of up to 50 mag if a steeper extinction law is assumed) and still
around 3 mag in the Ks-band \citep[e.g.][]{scoville2003,schoedel2010b}. The aligned interstellar dust grains responsible for the
polarization cause extinction as well, but non-aligned grains can also contribute. Therefore, the same particles are not
necessarily responsible for both effects \citep[e.g.][]{martin1990}. Universal power laws have been claimed for NIR extinction by
\cite{draine1989} and polarization \citep{martin1990}, who also showed that the law applicable to polarization in the optical domain
\citep{serkowski1975} is a poor approximation in the NIR. The power law indices presented in these works for the extinction
and the polarization power law are almost the same (1.5-2.0). It also appears that there is a correlation between the measured
extinction of an intrinsically unpolarized source and foreground polarization \citep{serkowski1975}, but this relation is quite
complex. In the light of new results for the extinction law, which seem to deviate consistently from the Draine law
\citep[e.g.][]{gosling2009,schoedel2010b}, new polarimetric measurements may be useful to further clarify the relation between
extinction and polarization.
\begin{figure}[!t]
\centering
\includegraphics[width=\textwidth,angle=0, scale=0.48]{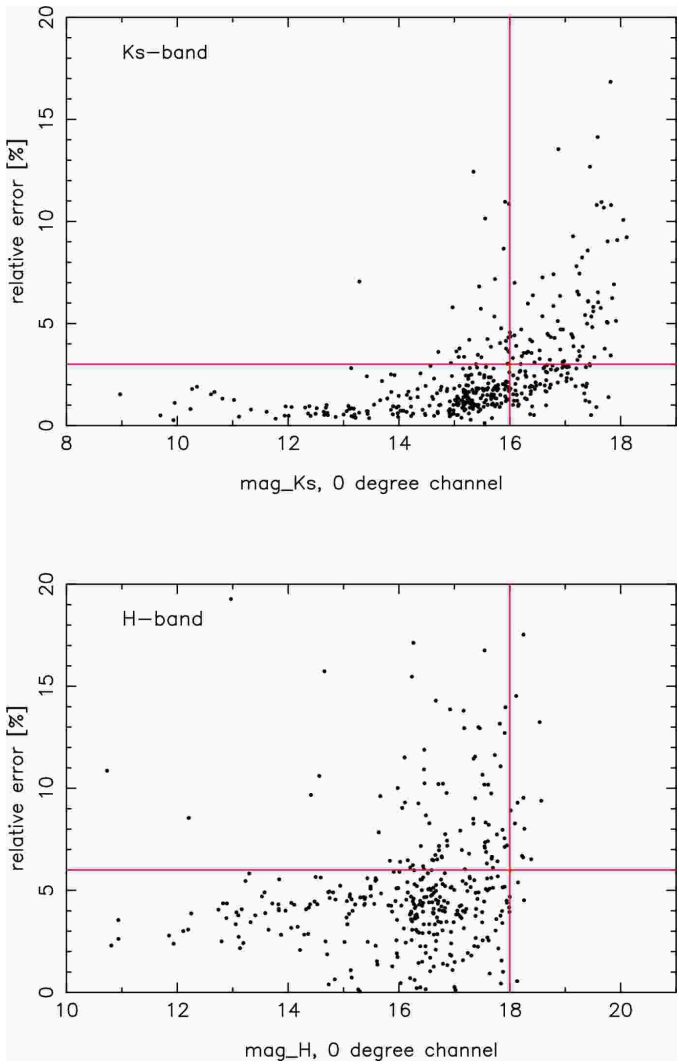}
\caption{\small Total relative flux errors of the sources detected in the Ks- (upper frame) resp. H-band (lower frame).
Only the 0$^{\circ}$ channel is shown here,
but the error distribution is very similar in the other three channels. The values are based on variations between the fluxes
measured for the same source at different dither positions. The red lines denote the 16/18 mag resp. 3/6\% upper limits (see
Appendix \ref{SectionErrors}).}
\label{FigPhoterrsHK}
\end{figure}
The central parsec of the GC is a well-suited but challenging environment to study this relation,
because it contains a huge number of sources that exhibit large variances in extinction \citep[1-2 mag, see][]{buchholz2009,schoedel2010b},
which is produced along a long line-of-sight by a great number of dust clouds with possibly different
grain alignment and composition. We are using a new Ks-band extinction map of the central parsec recently been presented by
\cite{schoedel2010b}.\\
The aims of this work are to present the first map of stellar H-
and Ks-band polarization of the central few
arcseconds, a highly crowded region with more than 10 sources per arcsec$^2$. In order to avoid crowding and to resolve the
sources individually, the resolution of an 8m class telescope is required. Especially in the H-band, no polarization
measurements exist so far for individual resolved sources in the central parsec. In addition, we will present the first
spatially resolved polarimetric measurements
on IRS~1W and IRS~21, also resolving the bow-shock structure of IRS~21 for the first time in the Ks-band. Furthermore, we
will conduct a variability analysis on the extended sources IRS~1W, 5, 10 and 21 in the H-, Ks- and L-band.\\
In \S \ref{SectObs} we discuss the photometric methods used and the calibration applied
to the data. Our results will be presented in \S \ref{SectResults}, followed by a summary and discussion of
their implications in \S \ref{SectDiscussion}.
\section{Observation and data reduction}
\label{SectObs}
\subsection{Observation}
The polarimetric datasets used here were obtained using the NAOS-CONICA (NACO) instrument at the ESO VLT unit telescope 4
on Paranal in June 2004 (H-band broadband filter, program 073.B-0084A, dataset 1, see Tab.\ref{TabObservations}) and May 2009
(Ks-band broadband filter, program 083.B-0031A, dataset 2). We also used several additional polarimetric Ks-band datasets
contained in the ESO archive (datasets 3-15)\footnote{Based on observations collected at the European Organization for
Astronomical Research in the Southern Hemisphere, Chile} that covered IRS~21 as well through a rotated field-of-view (FOV).
Please see Tab.\ref{TabObservations} for details on the individual observations. For our variability study of the bow-shock
sources we used NACO imaging data taken between June 2002 and May 2008 (H-, Ks- and L'-band data contained in the ESO archive).\\
\begin{figure}[!t]
\centering
\includegraphics[width=\textwidth,angle=-90, scale=0.32]{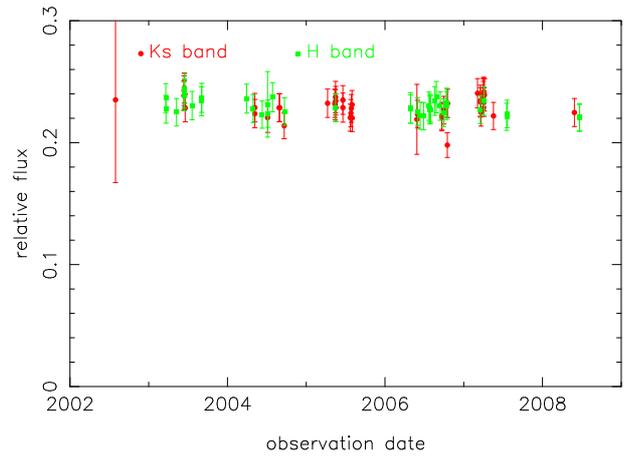}
\caption{\small H- (green) and Ks-band (red) lightcurve of the known non-variable source IRS~16C.}
\label{FigNonvar}
\end{figure}
The seeing during the polarimetric Ks-band observations was excellent with a value of $\sim$0.5'', while conditions
during the H-band observations were less optimal with a seeing of $\sim$0.8''. We used the bright super-giant IRS 7
located about 6'' north of Sgr A* to close the feedback loop of the adaptive optics (AO) system, thus making
use of the infrared wavefront sensor installed with NAOS. The sky background was determined by taking several
dithered exposures of a region largely devoid of stars, a dark cloud 713'' west and 400'' north of Sgr A*.
The Wollaston prism available with NACO in combination with a rotatable half-wave plate was used for the polarization
measurements. The two channels produced by the Wollaston prism (0$^{\circ}$ and 90$^{\circ}$), combined with two orientations
of the half-wave plate (0$^{\circ}$ and 22.5$^{\circ}$), yielded four sub-images for each of several dither positions along
the east-west axis. In total, we were able to cover a field-of-view of 3''$\times$16.5'' (Ks-band) respectively
3''$\times$19'' (H-band), corresponding to 0.12 pc $\times$ 0.66 pc and 0.12 pc $\times$ 0.72 pc, respectively (see
Fig.\ref{FigFOVwoll}).\\
All images were corrected for dead/hot pixels, sky-subtracted and flat-fielded. It is essential for a correct calibration that
the flat-field observations are taken through the Wollaston, because the flat-field shows variations caused
by different transmissivity of the channels as well as effects of the inclined mirrors behind the prism
\citep[see][for details on these effects]{witzel2010}. Using a flat-field taken without the Wollaston prism can produce
offsets in the order of 2\% and 10$^{\circ}$ in the measured polarization parameters.
Fig.\ref{FigFOVwoll} shows the region covered by the final mosaic that was used for the polarimetry.\\
\begin{figure*}[!t]
\centering
\includegraphics[width=\textwidth,angle=0, scale=1.0]{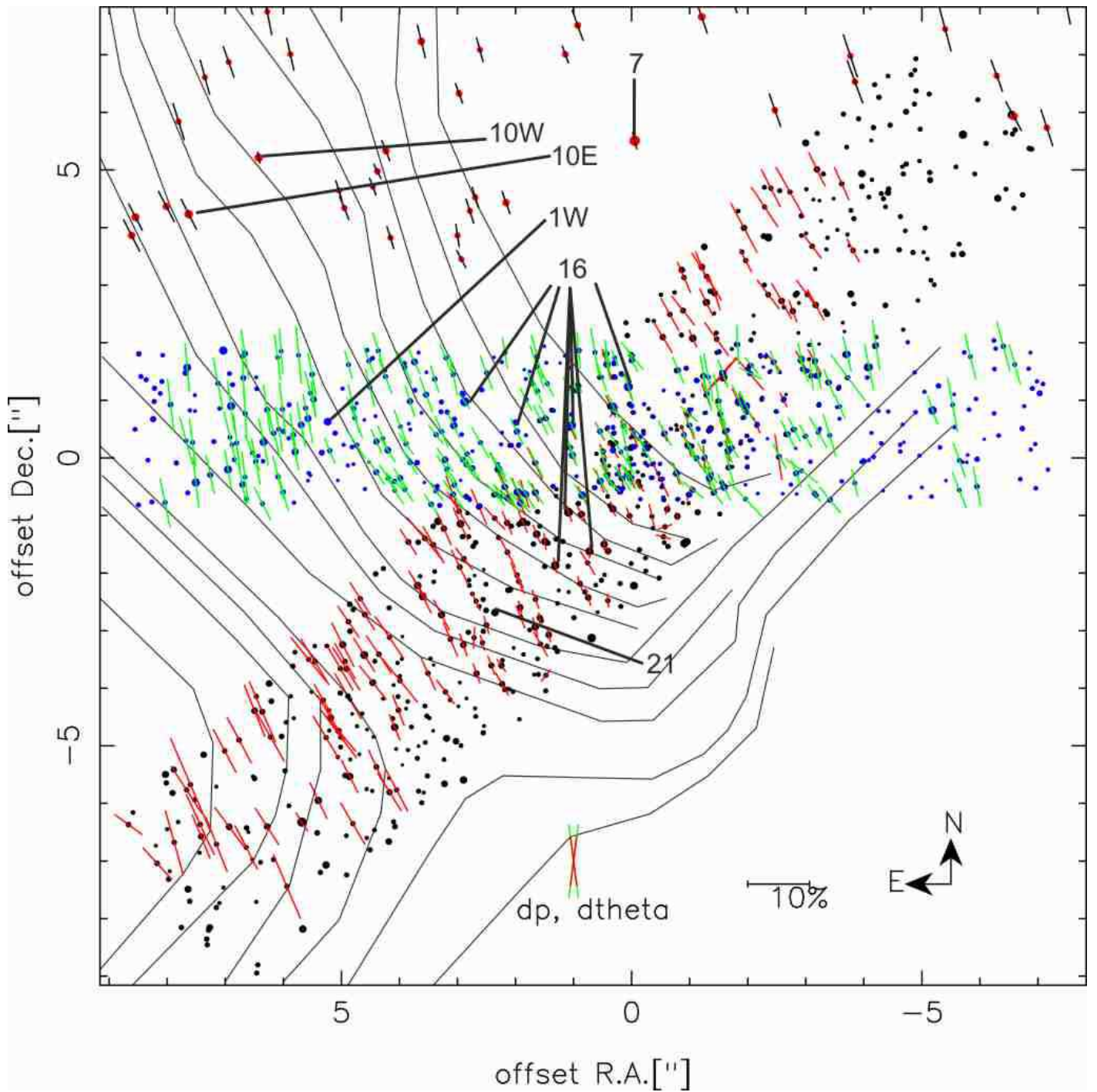}
\caption{\small Ks-band polarization map of stars in the Galactic center. Only reliably measured polarization values are shown
here. Blue circles and green lines: dataset 2. Black circles and red lines: dataset 4. The diameter of the circles corresponds
to the brightness of the source. For comparison, we also show preliminary results of new observations taken in March 2011 (denoted
by red circles and black lines). Typical errors are indicated by error cone (lower center). Thin black lines in the background denote magnetic fields
determined from MIR data \citep[][, based on a 1.5'' beam]{aitken1998}. The brightest sources are also indicated.}
\label{FigPolmapK}
\end{figure*} 
The data used here were originally taken to examine flares of Sgr A* \citep[see e.g.][]{eckart2006,meyer2006,zamaninasab2010}, and this purpose requires the highest possible
time resolution. This is the reason why differential polarimetry could not be applied. This technique can eliminate or reduce
many instrumental effects and thus increase precision by rotating the imager by 90$^{\circ}$ between exposures and thus canceling
these effects out (effectively switching the 0 and 90$^{\circ}$ channels). But this also means more time is needed for each
exposure, because two images have to be taken for one data-point and the actual rotating takes time as well. 
\begin{figure*}[!t]
\centering
\includegraphics[width=\textwidth,angle=-90, scale=0.64]{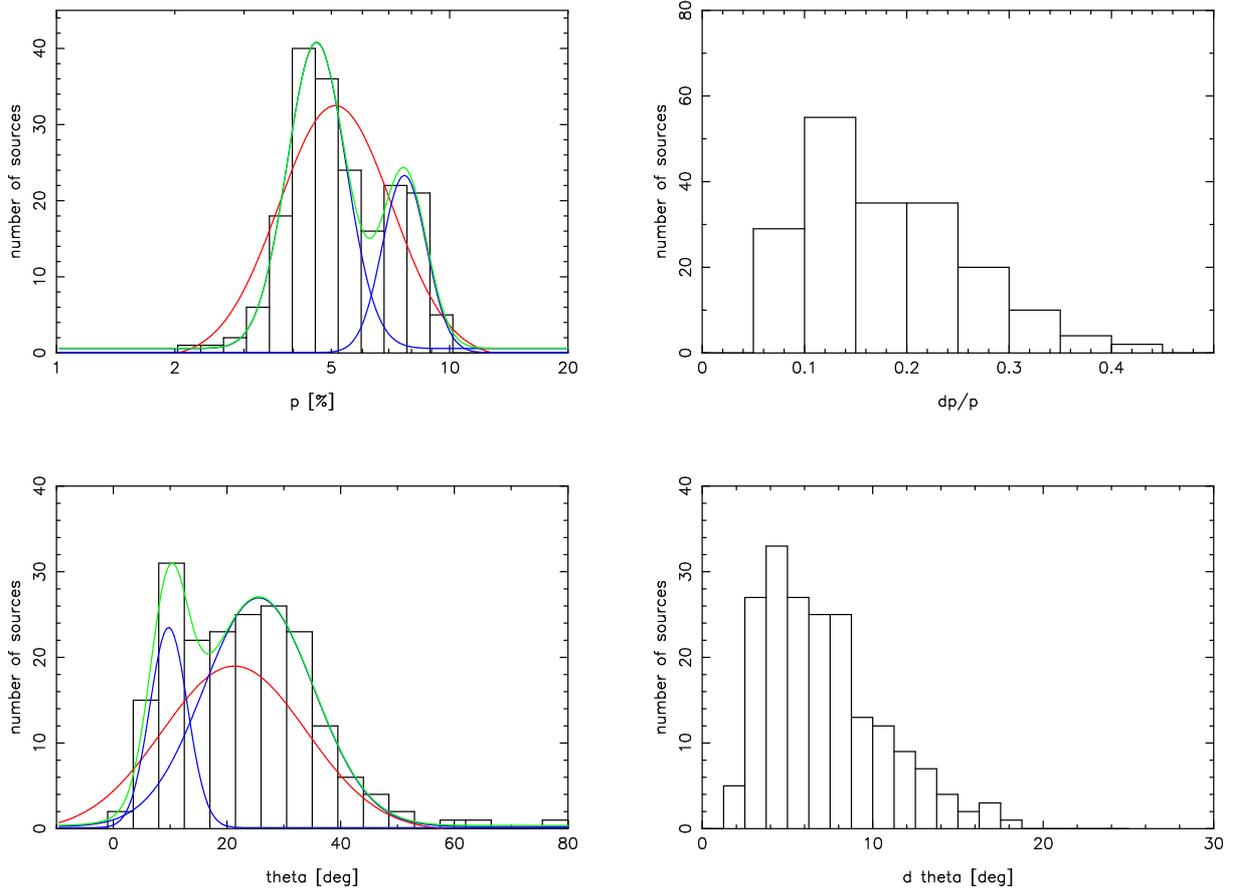}
\caption{\small Ks-band polarization degrees (plotted on logarithmic scale, upper left) and angles (lower left) of stars in
the Galactic Center (2009-05-18, dataset 2). The red line denotes the fit with one Gaussian distribution, while the green resp. blue lines denote
the fit with a double Gaussian (green: sum, blue: individual Gaussians).  Upper right: relative errors of the polarization degrees.
Lower right: absolute errors of the polarization angles.}
\label{FigPoldegsK}
\end{figure*}
\subsection{Photometry}
\subsubsection{Deconvolution-assisted large-scale photometry}
For all datasets, the individual exposures were combined to a mosaic. All photometry was conducted on these mosaics.\\
Accurate photometry is crucial for polarimetry, especially when the polarization of the targets is of just a few percent.
This is the case for the sources in the central parsec in the H- and Ks-band, and the effects of crowding and variations of the
point-spread-function (PSF) over the FOV complicate photometry even more. For the very bright sources in the FOV, such as the
IRS-16 and IRS-1 sources and the extended sources IRS-21 and IRS-1W, crowding is not a problem, but saturation can lead to
additional complications. These effects have to be countered effectively in order to achieve low photometric errors.\\
We therefore adopted a photometric method recently presented by \cite{schoedel2010a}: first, we used
the StarFinder IDL code \citep{diolaiti2000} to repair the cores of saturated sources (only necessary for
some very bright sources in the H-band image) and extract a PSF on the full image
from sufficiently bright and isolated sources. For this first step, the most suitable source would be the
guide star IRS~7 itself, since it is several magnitudes brighter than any source within several arcseconds,
but this source was not covered by the FOV of our dataset. We used a PSF determined from several IRS~16 and IRS~1 sources instead.
Unfortunately, the most suitable of these sources were only contained in the FOV of the Ks-band data, but not covered in the H-band.
Since this process is designed to determine the faint wings of the PSF accurately \citep[see][]{buchholz2009,schoedel2010a},
it works worse the fainter the initially used PSF sources are.\\
We used this PSF for a linear deconvolution, i.e. a division in Fourier space, followed by the application of a Wiener
filter. We then applied local PSF fitting photometry by using StarFinder on overlapping sub-frames of the deconvolved image
($\sim$3''$\times$10.6'', using wider sub-fields than \cite{schoedel2010a}, this generates large overlapping regions and
ensures that enough bright sources are contained in
each subfield for the PSF estimation). This method significantly reduces source confusion and also minimizes systematic errors introduced
by variations of the shape of the PSF over the FOV due to anisoplanasy. Despite the relatively narrow FOV, this effect still occurs
because the distance and the angle to the guide star change considerably over the field. While using narrower sub-fields might
counteract this, the availability of sufficiently bright PSF stars takes precedence, since this factor is the primary limit for the
quality of the photometry.
\\
The resulting fluxes in all four channels were normalized to an average of one over all channels for each source and
then merged to a common list of sources detected in all channels of each sub-frame. The sub-frame lists were then
merged to a common list of all detected sources.\\
Fig.\ref{FigPhoterrsHK} show the photometric uncertainties in the H- and Ks-band. We only used sources brighter
than 16 mag (Ks-band)
respectively 18 mag (H-band) and with relative photometric errors of less than 3\% resp. 6\% in the subsequent analysis. The lower
brightness limit was chosen to avoid problems with insufficient completeness and unreliable photometry, because the errors
increase drastically for fainter sources. Please see Appendix \ref{SectionErrors} for details on our error estimation and the reasons for the error threshold.
\begin{figure*}[!t]
\centering
\includegraphics[width=\textwidth,angle=-90, scale=0.64]{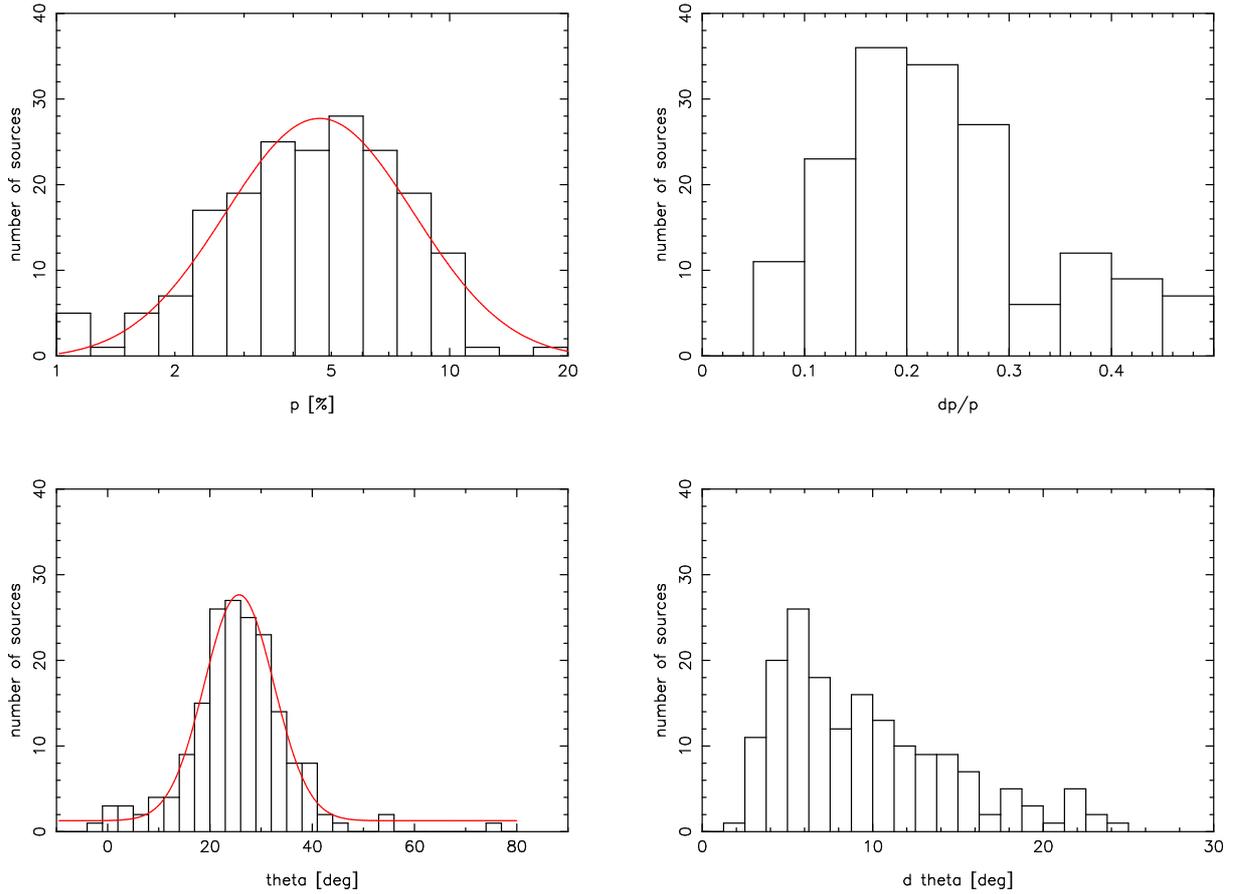}
\caption{\small Ks-band polarization degrees (plotted on logarithmic scale, upper left) and angles (lower left) of stars in
the Galactic Center (2007-04-03, dataset 4). The red lines denotes fits with a Gaussian distribution. Upper right: relative errors
of the polarization degrees. Lower right: absolute errors of the polarization angles.}
\label{FigPoldegsKtwisted}
\end{figure*}
\subsubsection{Photometry on extended sources}
Applying the PSF fitting algorithm to extended sources leaves large residua, and a simple core-subtraction
with a stellar PSF does not counter any distortions of the extended component produced by the atmosphere and the telescope.
But high precision is necessary here as well.
We therefore used a different approach: we deconvolved a small region of the mosaic images with the Lucy-Richardson
algorithm (the ''ringing'' produced by the linear deconvolution complicates the photometry of extended features), using the
PSF estimated from the complete mosaic. This resulted in a much clearer view of the extended features (see
Fig.\ref{FigIRS1W2dKs}), while other sources in the vicinity appear point-like.\\
We shifted the resulting images to a common reference frame. We then covered the extended features with overlapping apertures
($\sim$ 27 mas radius), measuring the flux in each aperture. From these fluxes we determined the polarization at the position
of the apertures. For comparison, the same method was applied to the PSF used for the deconvolution (see Appendix \ref{SectSimLR}).    
In addition, we determined the total polarization of the two known extended sources in our FOV, IRS~1W and IRS~21, by
covering them with apertures of 0.25'' radius. This allowed a comparison to previous observations.\\
In order to study a possible flux variability of the extended sources (see \S \ref{SectExtended}), we used 45 Ks-band and
38 H-band NACO datasets contained in the ESO archive (taken between 2003 and 2008). Where the bright point-sources were
saturated, we used StarFinder to
repair their cores. We then applied aperture photometry to the extended sources and several known bright non-variable sources
\citep[see][]{ott1999,rafelski2006}, with additional apertures placed on regions devoid of stars close to the targets
to obtain a background estimate. This background was subtracted from the recovered fluxes and the result normalized to
the total flux of the chosen non-variable calibration sources for each dataset \citep[the IRS~16 sources except 16SW and 16NE and
the IRS~33 sources; IRS~16NE is strongly saturated, while IRS~16SW has been described as an eclipsing binary and therefore shows
strong flux variabilities, making it unsuitable as a reference source, see][]{ott1999,rafelski2006}. We estimated the uncertainty
of each flux measurement by repeating the aperture
photometry on the individual non-mosaiced images of each dataset (which can be assumed to be independent measurements), and
adopted the standard deviation of the recovered fluxes as the total flux error. Especially for bright and isolated sources, this
yields very small statistical errors on the order of less than 0.5\%. We still find systematic variations of the measured fluxes
on the order of about 5\% between the epochs even for sources known to be non-variable on the timescale used here
(see Fig.\ref{FigNonvar}). The latter value provides a more realistic estimate of the photometric accuracy, so we introduced an
additional flux error of 5\% for all data-points to be able to separate real variability from noise.
\begin{figure*}[!t]
\centering
\includegraphics[width=\textwidth,angle=-90, scale=0.32]{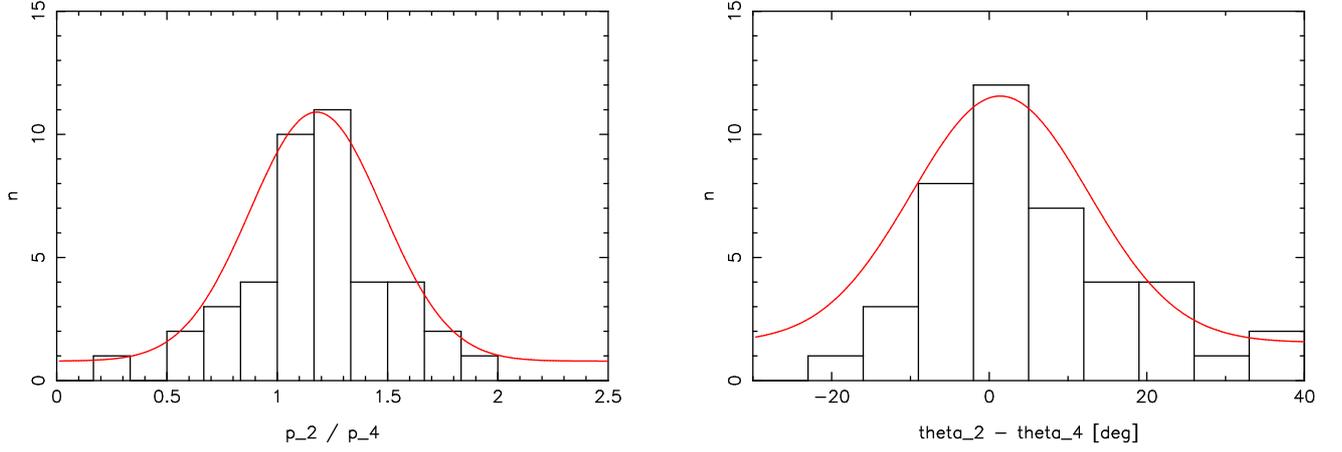}
\caption{\small Comparison between polarization degrees (left frame) and angles (right frame) measured on sources common to
dataset 2 (2009) and dataset 4 (2007). Plotted are histograms of $\frac{p_2}{p_4}$ resp. $\theta_2 - \theta_4$, with Gaussians
fitted to both values (red lines).}
\label{FigKpolcomp}
\end{figure*}
\subsection{Polarimetry}
We determined the polarization degree and angle of each source by converting the measured normalized fluxes
into normalized Stokes parameters:
\begin{eqnarray}
I &=& 1\nonumber\\
Q &=& \frac{f_{0}-f_{90}}{f_{0}+f_{90}}\\
U &=& \frac{f_{45}-f_{135}}{f_{45}+f_{135}}\\
V &=& 0.\nonumber
\end{eqnarray}
Because NACO is not equipped with a $\frac{\lambda}{4}$ plate, it was not possible to measure circular polarization.
However, \cite{bailey1984} showed that the circular polarization of sources in the GC is at best very small, so we
assume here that it can be neglected and set to 0 at our level of accuracy. Polarization degree and angle can then be determined as
\begin{eqnarray}
P &=& \sqrt{Q^2+U^2}\\
\theta &=& 0.5 \times atan \left( \frac{U}{Q} \right).
\end{eqnarray}
Errors for I, Q, U  and subsequently p and $\theta$ were determined from the flux errors. 
In order to determine whether or not the polarization of a source was determined reliably, we calculated the normalized
fluxes that would be expected for the determined values of
p and $\theta$. The difference between these values and the measured fluxes was then compared to the photometric errors
of each data-point. The source was only classified as reliable if the root-mean-square of the deviations did not exceed
the root-mean-square of the relative photometric errors. The subsequent analysis is only based on these higher quality
sources.
\subsection{Calibration of the measured polarization}
\label{SectCalibration}
A first comparison of our measurements to known values \citep{knacke1977,bailey1984,eckart1995,ott1999}
revealed significant offsets, especially in the polarization angles. Instead of the expected orientation along
the galactic plane (oriented $\sim$31.4$^{\circ}$ East-of-North, in the following, positive
angles should be read as East-of-North, negative as West-of-North), we found orientations of the polarization
vectors of about -5$^{\circ}$. The problem that surfaces here is that NACO was not specifically built for polarimetry,
so instrumental effects like this can be expected and have to be countered by a special calibration. Also, the observation
technique was developed for the study of short-term variabilities (flares) of Sgr A* and not for high-precision polarimetry
on stellar sources.\\
To reduce this problem, \cite{witzel2010} developed an analytical model of the behavior of polarized light within NACO.
It consists of M\"uller matrices to be applied to the measured Stokes vector of each source, which then
yields the actual Stokes parameters of the source. In general, any optical effect such as reflection, transmission,
polarization etc. can be described by a M\"uller matrix. A combination of effects as it occurs here is then represented
by a multiplication of the individual M\"uller matrices. If the necessary material constants and the construction of
the instrument are known, the resulting matrix can be used to significantly reduce systematic offsets and uncertainties.
\cite{witzel2010} show that by using this method, the systematic uncertainties of polarization degrees and angles can
be reduced to $\sim$1\% and $\sim$5$^{\circ}$. Applied to our data, this limits the final accuracy by the photometric 
uncertainties instead of by instrumental effects. For a detailed description of the model and the M\"uller
matrices themselves, see \cite{witzel2010}.\\
Utilizing this new calibration model means that an actual direct calibration can be achieved for the first time at this
resolution. Previous studies like \cite{eckart1995} and \cite{ott1999} had to adopt a calibration based on reference values taken from
\cite{knacke1977}.\\
\subsection{Correcting for foreground polarization}
\label{SectForegroundremoval}
It can be assumed that the total effect of the foreground polarization can be treated as a simple linear polarizer with
a certain orientation $\theta_{fg}$ and efficiency $p_{fg}$. This can be described by a M\"uller matrix:
\begin{figure*}[!t]
\centering
\includegraphics[width=\textwidth,angle=-90, scale=0.30]{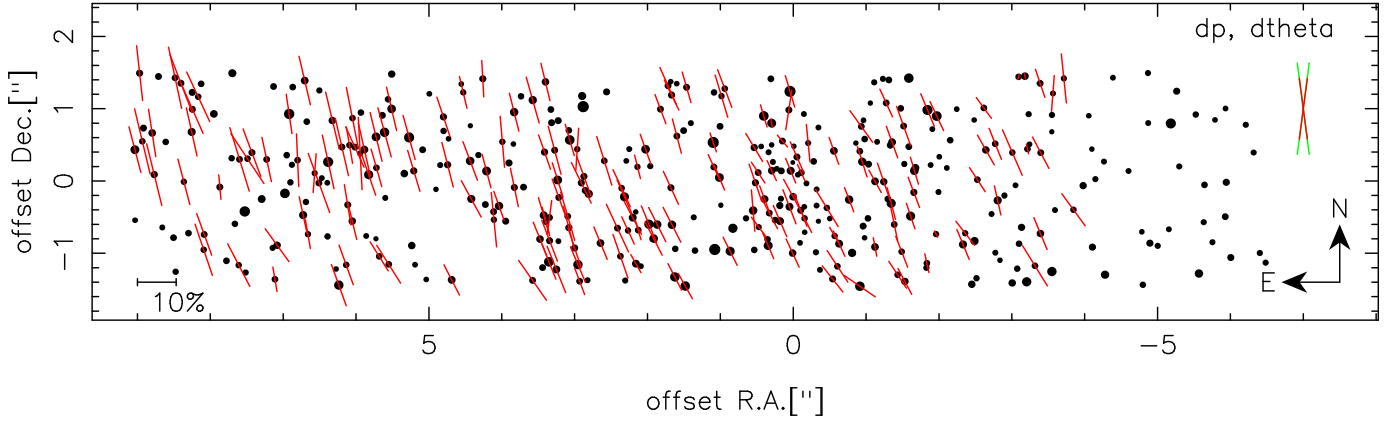}
\caption{\small H-band polarization map of stars in the Galactic Center (dataset 1). Only reliably measured polarization values
are shown here. Typical errors plotted as error cone (upper right), adopted from Ks-band data.}
\label{FigPolmapH}
\end{figure*}
\begin{eqnarray} S_{obs} &=& \left( \begin{array}{c}
I \\
Q \\
U \\
V \end{array} \right) = M_{fg} \times S_{int} = M_{fg} \times \left( \begin{array}{c}
I' \\
Q' \\
U' \\
V' \end{array} \right) \nonumber \\
&=& M_{rot}( - \theta') \times M_{lin}(p) \times M_{rot}(\theta') \times S_{int},
\end{eqnarray}
with $S_{obs}$ as the observed total Stokes vector and $S_{int}$ as the Stokes vector of the intrinsic polarization.
$M_{lin}(p)$ is the M\"uller matrix describing a linear polarizer, producing a maximum of polarization along the North-South-axis:
\[ M_{lin}(p) = \left( \begin{array}{cccc}
1 & -p & 0 & 0 \\
-p & 1 & 0 & 0 \\
0 & 0 & \sqrt{1-p^2} & 0 \\
0 & 0 & 0 & \sqrt{1-p^2} \end{array} \right) \].
This matrix has to be rotated to the appropriate angle by multiplying it with $M_{rot}(\theta')$, a standard 4$\times$4 rotation
matrix:
\[ M_{rot}(\theta') = \left( \begin{array}{cccc}
1 & 0 & 0 & 0 \\
0 & cos(2 \theta') & sin(2 \theta') & 0 \\
0 & -sin(2 \theta') & cos(2 \theta') & 0 \\
0 & 0 & 0 & 1 \end{array} \right) \].
$\theta' = 90^{\circ} + \theta_{fg}$ has to be used in the rotation matrix, because we define the polarization angle $\theta$ as
the angle where we measure the flux maximum, while the angle of reference for the M\"uller
matrix describing the linear polarizer is the angle where the maximum in absorption occurs.\\
For each source to which the depolarization matrix was applied, we used the average of the polarization parameters of the
surrounding point sources as an estimate for $\theta_{fg}$ and $p_{fg}$.\\ 
The resulting matrix can then be inverted and multiplied with the calibrated observed Stokes vector of a source to
remove the foreground polarization and leave only the intrinsic polarization. We applied this method to the total
polarization and the polarization maps of the extended sources (IRS~21 and IRS~1W), to isolate their intrinsic
polarization pattern. The relevance of the results depends heavily on the accuracy of the foreground polarization estimate. 
\section{Results and discussion}
\label{SectResults}
\subsection{Ks-band polarization}
\label{SectKpol}
\begin{figure*}[!t]
\centering
\includegraphics[width=\textwidth,angle=0, scale=1.0]{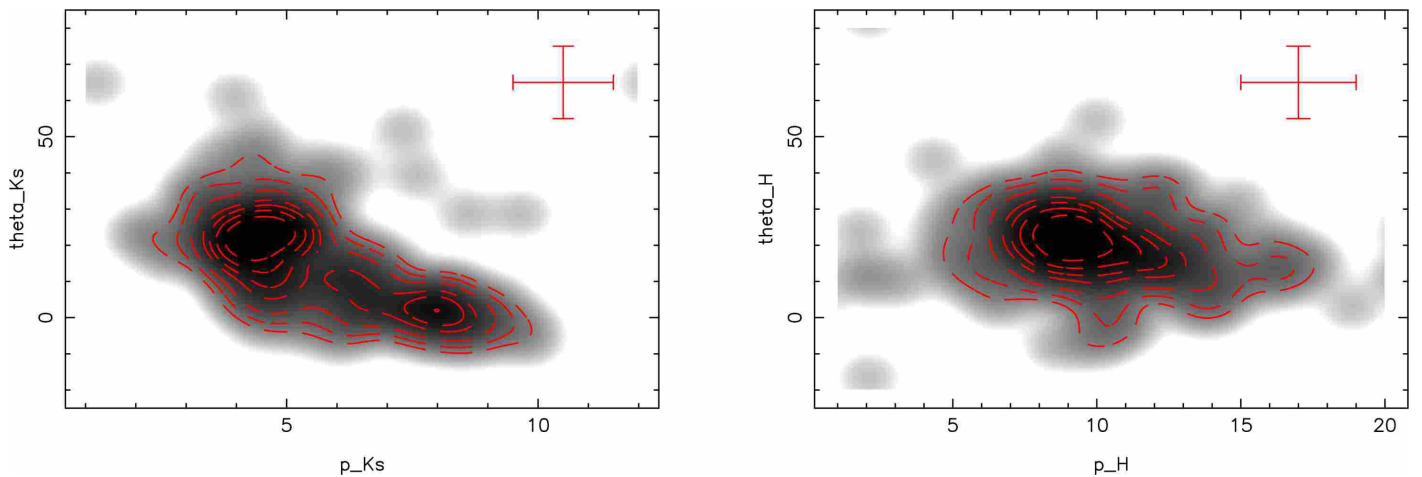}
\caption{\small Polarization angle vs. polarization degree, plotted as logarithmic smoothed point density, with typical error indicated
by red bars (upper right corners). Contours were plotted to guide the eye. Left frame: Ks-band, right frame: H-band}
\label{FigPolparsCompHK}
\end{figure*}

\paragraph{2009 data (dataset 2)}
We were able to measure reliable polarization parameters for 194 sources brighter than 16 mag in our main 2009 Ks-band dataset.
For fainter sources, the photometric uncertainty becomes too large to determine the polarization reliably, and this limit also
helps to avoid problems with insufficient completeness and source crowding, which could generate a bias in the averaged values. The
subsequent analysis and the comparison to the H-band data is only based on this dataset.\\
The polarization angles in the central arcseconds mostly follow the orientation of the galactic plane within the
uncertainty limits (31.4$^{\circ}$ \citep{reid2004}, while we find angles of $\sim$25-30$^{\circ}$ for our sources). Toward
the eastern edge of the FOV, we find slightly steeper angles ($\sim$5-15$^{\circ}$, see Fig.\ref{FigPolmapK}). A few
sources west of Sgr A* also show similar steep angles, but there are too few reliable sources there to allow
any conclusions. Here and in the following, we use the term ''steeper angle'' to denote angles with an absolute value
closer to 0$^{\circ}$, so that an angle of 10$^{\circ}$ would be steeper than one of 30$^{\circ}$. The polarization angle
can vary between -90$^{\circ}$ and 90$^{\circ}$, in the way that 91$^{\circ}$ (East-of-North) corresponds to -89$^{\circ}$.\\
The distribution of the polarization angles (see Fig.\ref{FigPoldegsK}, lower left frame) can be fitted with a single Gaussian,
peaked at 20$^{\circ}$ with a FWHM of 30$^{\circ}$. Using the FWHM as a measure for the uncertainty (with $\sigma =
\frac{FWHM}{2 \sqrt{2 \times ln(2)}}$), this yields $\theta = 20^{\circ} \pm 13^{\circ}$. Using a fitting function with two
Gaussian peaks yields a significantly lower $\chi^2$ (by a factor of 3). The two peaks are fitted at $\theta_1 = 10^{\circ} \pm
4^{\circ}$ (FWHM of 10$^{\circ}$) respectively $\theta_2 = 26^{\circ} \pm 8^{\circ}$ (FWHM of 20$^{\circ}$). Considering the
uncertainties of the polarization angles shown in the lower right frame of Fig.\ref{FigPoldegsK}, which are on the order of up
to 15$^{\circ}$, it can be questioned if these two peaks are indeed a real feature, with the distance between the peaks on the
order of these typical errors.\\
The polarization degree also appears to vary over the field, with values of 4-5\% in the central region and 8-10\% toward the
eastern edge (and for some western sources, but with the same caveat as for the polarization angle). Especially sources in the
area around the IRS~1 sources show these higher polarization degrees. We fitted the logarithms of the polarization degrees
with a Gaussian (peaked at (5.1 $\pm$ 1.7)\%, FWHM of 4.0\%, see Fig.\ref{FigPoldegsK}, upper left frame), and like the single Gaussian fit to the polarization angles, the fit was quite poor. Repeating the fit with a double Gaussian yielded two peaks at
(4.6 $\pm$ 0.8)\% respectively (7.7 $\pm$ 1.2)\%, with FWHMs of 1.8\% resp. 2.8\% and a significantly better $\chi^2$ (by a
factor of 8). The relative uncertainties of the polarization degree are on the order of
up to 30\% (see Fig.\ref{FigPoldegsK}, upper right frame), and this limits the confidence in the two fitted peaks.\\
Comparing the two fitted Gaussian distributions for both parameters, we find a similar number of stars contained
in the 10$^{\circ}$ and the 7.7\% peak ($\sim$25-30\%), resp. the 26$^{\circ}$ and 4.6\% peak ($\sim$70-75\%). This confirms
the general trends found in Fig.\ref{FigPolmapK} and indicates that the fitted peaks indeed correspond to a real feature.\\
We plotted the Ks-band polarization degrees versus the polarization angles in Fig.\ref{FigPolparsCompHK}, left frame, and there
appears to be a trend that higher polarization degrees coincide with steeper polarization angles, despite the large errors.
\begin{figure*}[!t]
\centering
\includegraphics[width=\textwidth,angle=-90, scale=0.64]{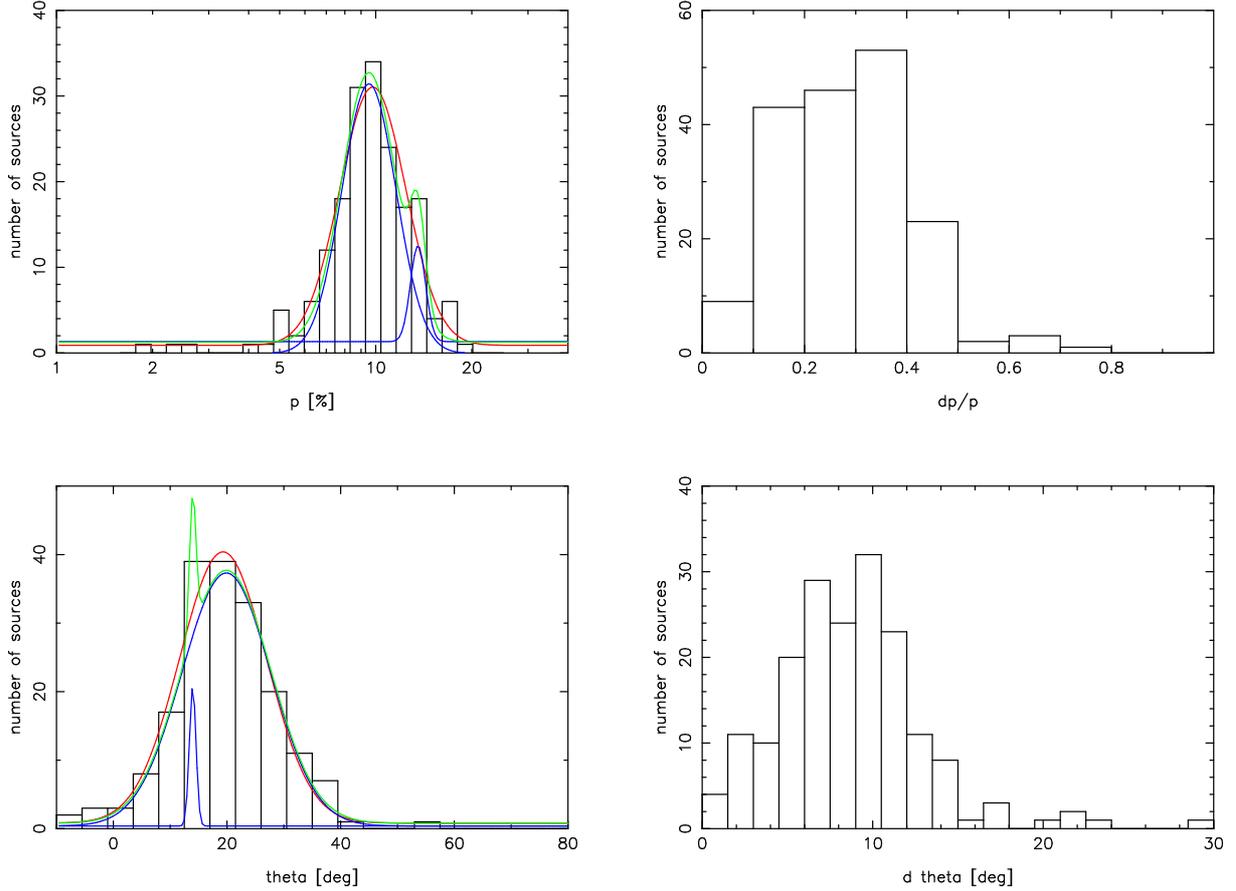}
\caption{\small H-band polarization degrees (plotted on logarithmic scale, upper left) and angles (lower left) of stars in
the Galactic Center (dataset 1). The red line denotes the fit with one Gaussian distribution, while green resp. blue lines denote the
fit with a double Gaussian (green: sum, blue: individual Gaussians). Upper right: relative errors
of the polarization degrees. Lower right: absolute errors of the polarization angles.}
\label{FigPoldegsH}
\end{figure*}
\paragraph{2007 data, rotated FOV (dataset 4)}
We measured reliable polarization parameters for 186 sources brighter than 16 mag. This dataset has a lower Strehl ratio (on average)
than the 2009 data (22\% compared to 27\%). We used this dataset only for comparison with the main Ks-band data (dataset 2, with the
main purpose to determine the presence of unaccounted instrumental effects over the FOV), since no H-band data with this FOV is
available.\\
The trends we find here are similar to those in the main dataset: the polarization angles appear to be aligned with the Galactic plane
(see Fig.\ref{FigPolmapK}), but we do not observe the same shift in polarization angle toward the east of the FOV. We do, however,
find an increase in polarization degree toward the southeast, similar to the increase found in the 2009 data toward the east.
Both the distribution of the polarization degrees (logarithmic) and the polarization angles can be fitted with a single Gaussian
with sufficient accuracy (see Fig.\ref{FigPoldegsKtwisted}), with peaks fitted at (4.6 $\pm$ 2.1)\% respectively $26^{\circ} \pm
8^{\circ}$ (FWHM of 5\% resp. 20$^{\circ}$). The FWHM of the distribution of the polarization degrees is comparable to that found
for the 2009 data, but a single Gaussian provides a much better fit here. The larger uncertainties would however lead to a blurring
of the two Gaussians, if indeed two were present.\\
The relative errors of the polarization degree (see Fig.\ref{FigPoldegsK}, upper right frame) mostly stay below 30\%, with some
outlier values of up to 40-50\%. This exceeds the errors found for dataset 2, but this can be expected because of the lower data
quality. The errors found for the polarization angles are also larger on average than those measured for dataset 2.\\
The sources in the overlapping area of dataset 2 and 4 for which reliable measurements could be obtained in both datasets show
very similar polarization parameters (see Fig.\ref{FigKpolcomp}). We find that the polarization degrees and angles of 82\%
of the common sources agree within 1 sigma, viewing the parameters individually. Both parameters agree within 1 sigma for 69\%
of the common sources. For 93\% of the common sources both parameters agree within 2 sigma, and 98\% (all but one source)
show an agreement within 3 sigma.\\
We fitted the relation between the polarization degrees resp. the difference of the polarization angles of the sources in the common
region of both datasets with a Gaussian (see Fig.\ref{FigKpolcomp}). We fitted the peaks at $\frac{p_2}{p_4} = 1.2 \pm 0.2$ (FWHM of
0.5) and $\theta_2 - \theta_4 = 1^{\circ} \pm 7^{\circ}$ (FWHM of $17^{\circ}$). The uncertainties of these values provide an estimate
for the general accuracy of the Ks-band measurements, so we plotted them as an error cone in Fig.\ref{FigPolmapK}.\\
This reinforces the confidence in the measured values in both datasets, while it also provides a confidence limit. A
$3 \sigma$-deviation would correspond to a (relative) 60\% offset in polarization degree and a difference of 21$^{\circ}$ in
polarization angle.\\
Unfortunately, the overlapping part of the FOVs is only
about 8 arcsec$^2$, and no data with a different FOV exist for the regions that show an excess in polarization degree/angle in
both datasets.\\ 
Fig.\ref{FigPolmapK} also shows the preliminary results of new polarization measurements (program 086.C-0049(A)) of a region north
of the FOV of dataset 2. These measurements seem to link up well with the findings in dataset 2 and 4. A detailed analysis of
these observations will be the subject of a future study.\\
\subsection{H-band polarization}
\label{SectHpol}
Reliable results could be obtained for 163 sources brighter than 18 mag. The limit of 18 mag was chosen because this corresponds to
the limit of 16 mag in the Ks-band, assuming a typical H-Ks of $\sim$2 \citep[with H-Ks$_{intrinsic} \sim$0 and A$_H$-A$_{Ks} \sim$2,
see e.g.][]{schoedel2010b}. The lower number
of sources with reliable polarization compared to the Ks-band can be attributed to the significantly lower Strehl ratio of the H-band
data (0.17 compared to 0.27 in the Ks-band) and the slightly different FOV, but the latter is a minor effect. We find polarization
angles very similar to those in the Ks-band, but with a more uniform distribution over the FOV (see Fig.\ref{FigPolmapH}). The
distribution of the polarization angles can be fitted well with a single Gaussian, peaking at $20^{\circ} \pm 8^{\circ}$ (FWHM of 
19$^{\circ}$, see Fig.\ref{FigPoldegsH}, lower left frame). Fitting this distribution with a double Gaussian produces a slightly
better $\chi^2$, but this can be expected for an increase in the number of fitting parameters. A single Gaussian fits the distribution
with sufficient accuracy, compared to the poor fit with a single Gaussian function for the Ks-band polarization angles. We find
typical errors of the polarization angle of up to 12$^{\circ}$ (see Fig.\ref{FigPoldegsH}, lower right frame).\\
The polarization degree also appears to be quite uniform over the FOV, with typical values of 8-12\%. Fitting the logarithms of the
polarization degrees with a single Gaussian leads to a peak at (9.8 $\pm$ 0.7)\% (FWHM of 1.7\%), satisfyingly matching the data (see
Fig.\ref{FigPoldegsH}, upper left frame). We also fitted the data with two Gaussian peaks for comparison, but this only marginally
improves the fit. We find relative uncertainties of the polarization degree on the
order of up to 40\% (see Fig.\ref{FigPoldegsH}, upper right frame).\\
Unfortunately, no high-resolution H-band measurements with a different FOV are available for a comparison.\\ 
As for the Ks-band, we plotted the H-band polarization degrees versus the polarization angles (see Fig.\ref{FigPolparsCompHK}, right
frame). No large-scale trend is visible, both parameters appear concentrated around $\sim$20$^{\circ}$ resp. $\sim$9\%.
\subsection{Comparison to previous results}
\label{SectComparisonPrev}
We compared the polarization degrees and angles of 30 sources from \cite{eckart1995} and 13 sources from \cite{ott1999}
to the values we determined from our data. These two studies were calibrated based on polarization parameters determined
by \cite{knacke1977} and \cite{lebofsky1982}. We find that 87\% of the Eckart sources agree with our values within
3 sigma in polarization degree (and 83\% in polarization angle). For the Ott sources (with a slightly different FOV)
we find that 77\% agree with our own values within 3 sigma for both polarization degree and angle. Differences between
these older studies and our own measurements can probably be attributed to the lower spatial resolution of the former.
Both older studies show average polarization angles generally parallel to the Galactic plane, on average at 25$^{\circ}$ resp. 
30$^{\circ}$. We find similar results for our much larger number of sources. The average polarization degree is slightly higher,
but this can be expected because the inclusion of IRS~7 with its polarization degree of only 3.6\% in both older surveys  
considerably lowered the flux-weighted average that was calculated there.\\
One other thing has to be considered here: we have for the first time applied an absolute polarimetric 
calibration to high angular resolution data. That we find such good agreement with data where a relative calibration based on
the \cite{knacke1977} values was used increases the confidence in both this study and the results of the mentioned previous works.\\
The only direct possibility to compare our study with the results of \cite{knacke1977} exists in the vicinity of IRS~1. There,
we find a flux-weighted average polarization of $4.3 \pm 0.5$\% at $14^{\circ} \pm 10^{\circ}$ in a 3.5'' aperture, which matches the value given in the older study ($3.5 \pm 0.5$\%
at $16^{\circ} \pm 5^{\circ}$). We calculated the flux-weighted average by summing the fluxes of the individual stars contained
in the aperture for each channel and calculating Q and U (and subsequently p and $\theta$) from these total values. This corresponds
to a flux-weighted average over Q and U.\\
The reason why this value is much lower than the polarization generally found in this area is the contribution of IRS~1W with its
high flux and an intrinsic polarization which is almost perpendicular to that of the sources in the vicinity.\\
This additionally supports our findings of higher polarization degrees toward the eastern edge of the FOV: for the total
polarization to be on the order of 4\% (including IRS~1W), the surrounding sources must have a significantly higher
polarization degree.\\
We compared our findings to several older NACO data-sets with the same FOV. Using aperture photometry on the IRS~1 sources
(except IRS~1W) and the northern IRS~16 sources except IRS~16NW (to avoid problems with saturation), we found average offsets
in both polarization degree and angle in the order of 10-15$^{\circ}$ and 2-3\%, with higher polarization degrees and steeper
polarization angles for the IRS~1 sources. This again confirmed the trends we find in our main data-set.\\   
Very few H-band polarization measurements are available for a comparison. The most recent survey with an aperture not
exceeding our FOV was conducted by \cite{bailey1984} in the J-, H- and K-band, who present polarization parameters
for two sources in our FOV, IRS~1 and IRS~16 (treating these complexes as a single source each, using a 3.0'' aperture).
That study measured $9.9 \pm 0.6$\% at $20^{\circ} \pm 1^{\circ}$ for IRS~1 respectively $10.2 \pm 0.2$\% at $8^{\circ}
\pm 1^{\circ}$ for IRS~16 in the H-band. Calculating a flux weighted average in a 3.0'' aperture around the IRS~1 resp.
IRS~16 sources based on our data yields values of $10.9 \pm 0.5$\% at $15^{\circ} \pm 10^{\circ}$ for IRS~1 and $7.2 \pm 0.5$\%
at $21^{\circ} \pm 10^{\circ}$ for IRS~16. This agrees well for IRS~1, while the polarization degree found for IRS~16 deviates
considerably. But it has to be considered that both sources were only very poorly resolved at the time of that study, not
all IRS~16 sources are contained in our FOV, and we only used sources with reliably measured polarization for the comparison.\\  
Our measured polarization degrees and angles are also compatible to the larger-scale polarization maps presented by
\cite{nishiyama2009} in the H- and the Ks-band. The authors find polarization angles of $\sim$20$^{\circ}$ in the area around our own
FOV, which itself is not covered in that study.
\begin{figure*}[!t]
\centering
\includegraphics[width=\textwidth,angle=-90, scale=0.34]{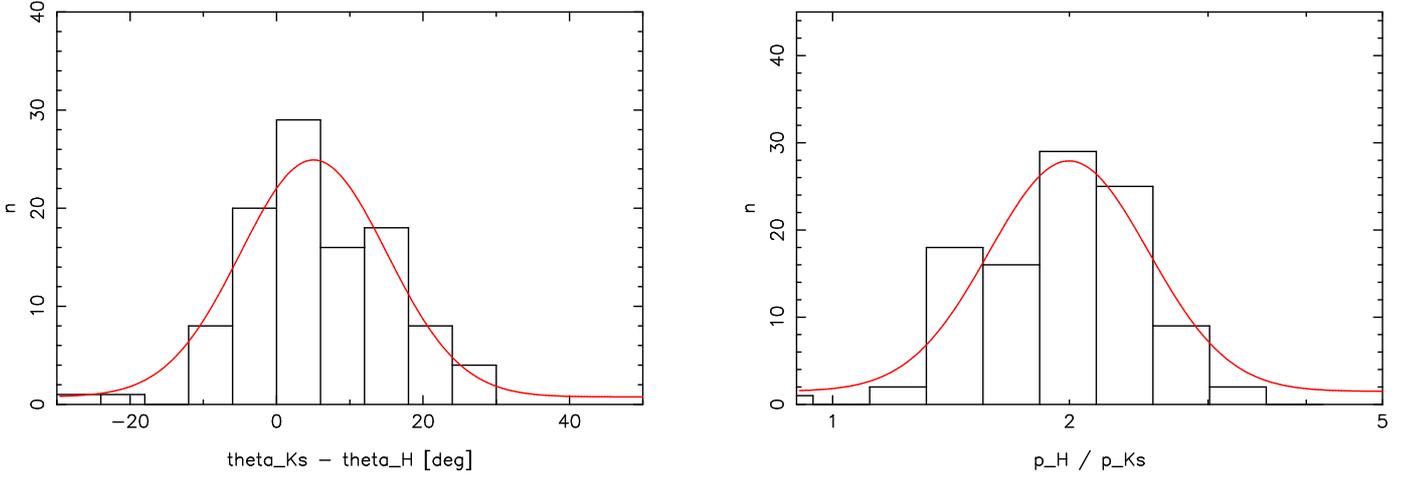}
\caption{\small Left frame: difference between H- and Ks-band polarization angle. Right frame: relation of H- to Ks-band polarization
degree (logarithmic plot). The red line represents a Gaussian fitted to the histograms.}
\label{FigPolcompHKtot}
\end{figure*}
\subsection{Relation between H- and Ks-band polarization}
\label{SectHKrelation}
\begin{figure*}[!t]
\centering
\includegraphics[width=\textwidth,angle=-90, scale=0.48]{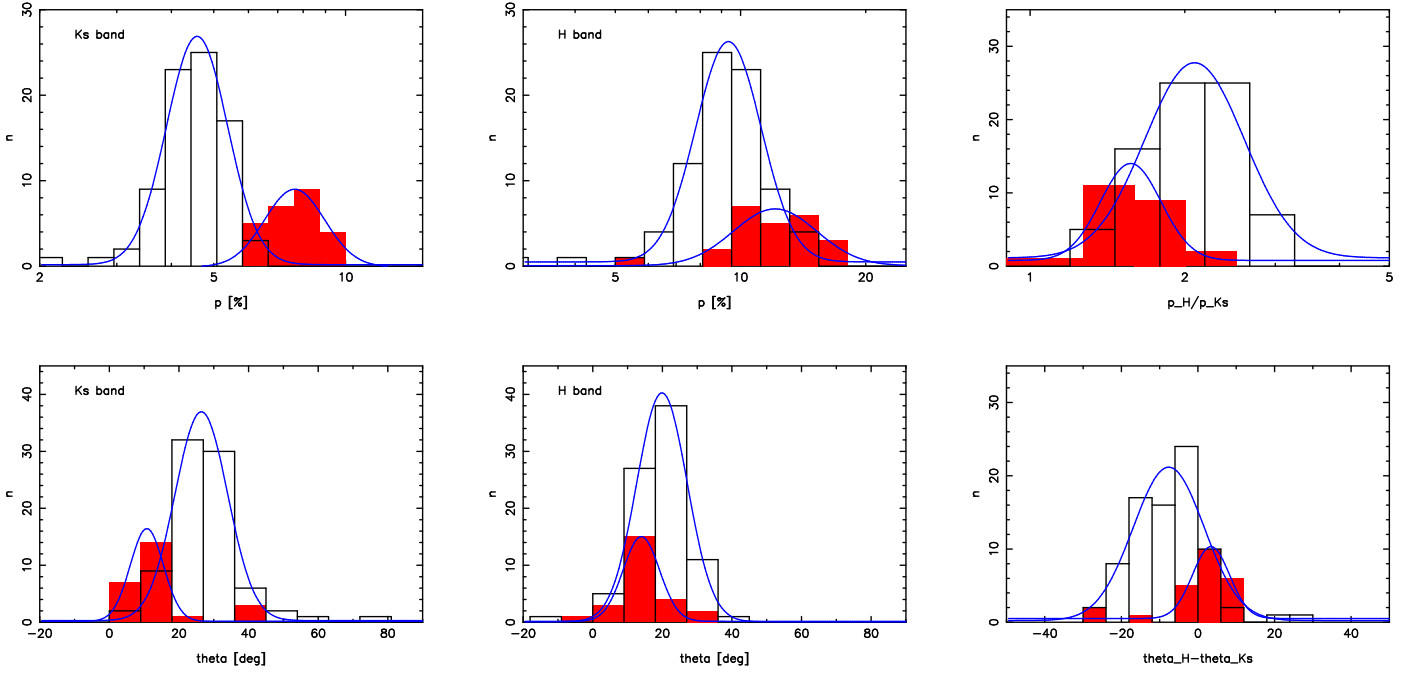}
\caption{\small Sources detected in H- and Ks-band, separated by Ks-band polarization. White columns: p$_{Ks} < 6$\%. Red columns:
p$_{Ks} > 6$\%. Upper/lower left frame: Ks-band polarization degree/angle, Upper/lower middle frame: H-band polarization degree/angle, 
upper right frame: relation of H- and Ks-band polarization degree (logarithmic plot). Lower right frame: difference between H- and
Ks-band polarization angle. Blue lines represent Gaussian fits to the histograms.}
\label{FigPolcompHKpolsep}
\end{figure*}

\begin{figure*}[!t]
\centering
\includegraphics[width=\textwidth,angle=-90, scale=0.48]{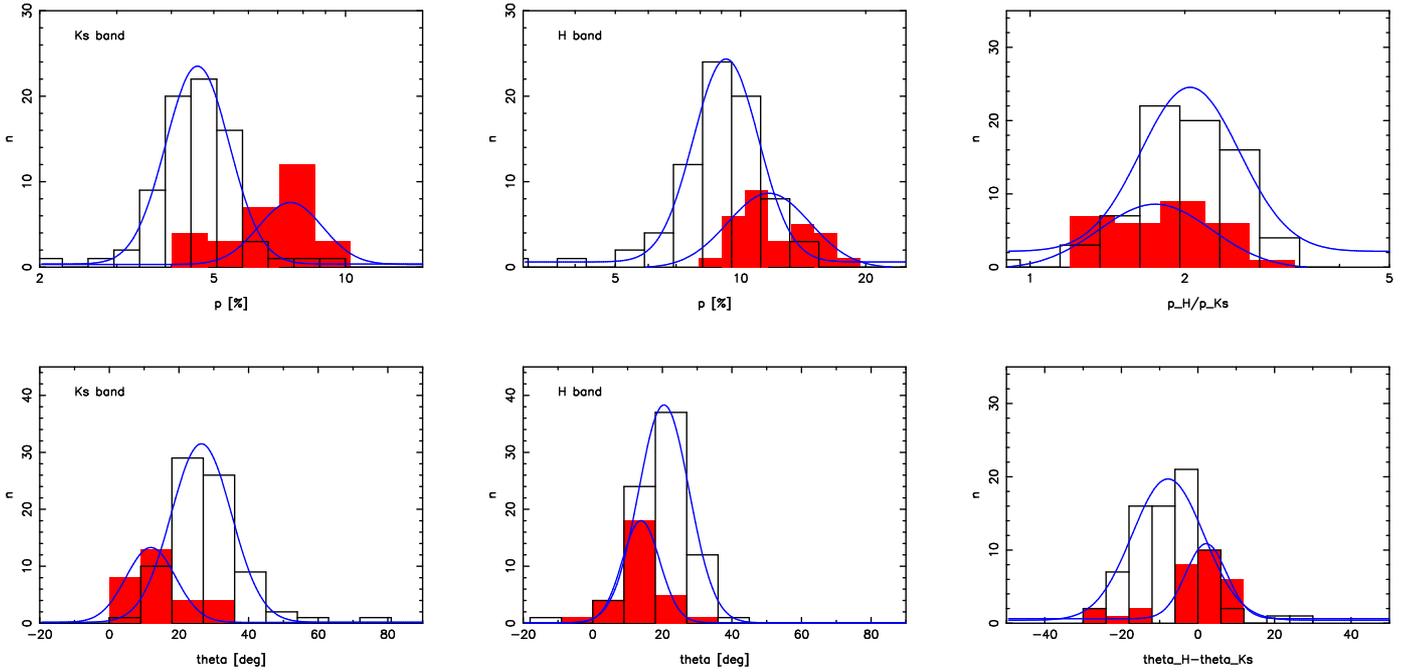}
\caption{\small Sources detected in the H- and Ks-band, separated by position along the East-West-axis. White columns: sources less
than 4.1'' east of Sgr A*. Red columns: sources more than 4.1'' east of Sgr A*. Upper/lower left frame: Ks-band polarization
degree/angle, upper/lower middle frame: H-band polarization degree/angle, Upper right frame: relation of H- and Ks-band polarization
degree (logarithmic plot). Lower right frame: difference between H- and Ks-band polarization angle. Blue lines represent Gaussian fits
to the histograms.}
\label{FigPolcompHKxpos}
\end{figure*}

\begin{figure*}[!t]
\centering
\includegraphics[width=\textwidth,angle=-90, scale=0.7]{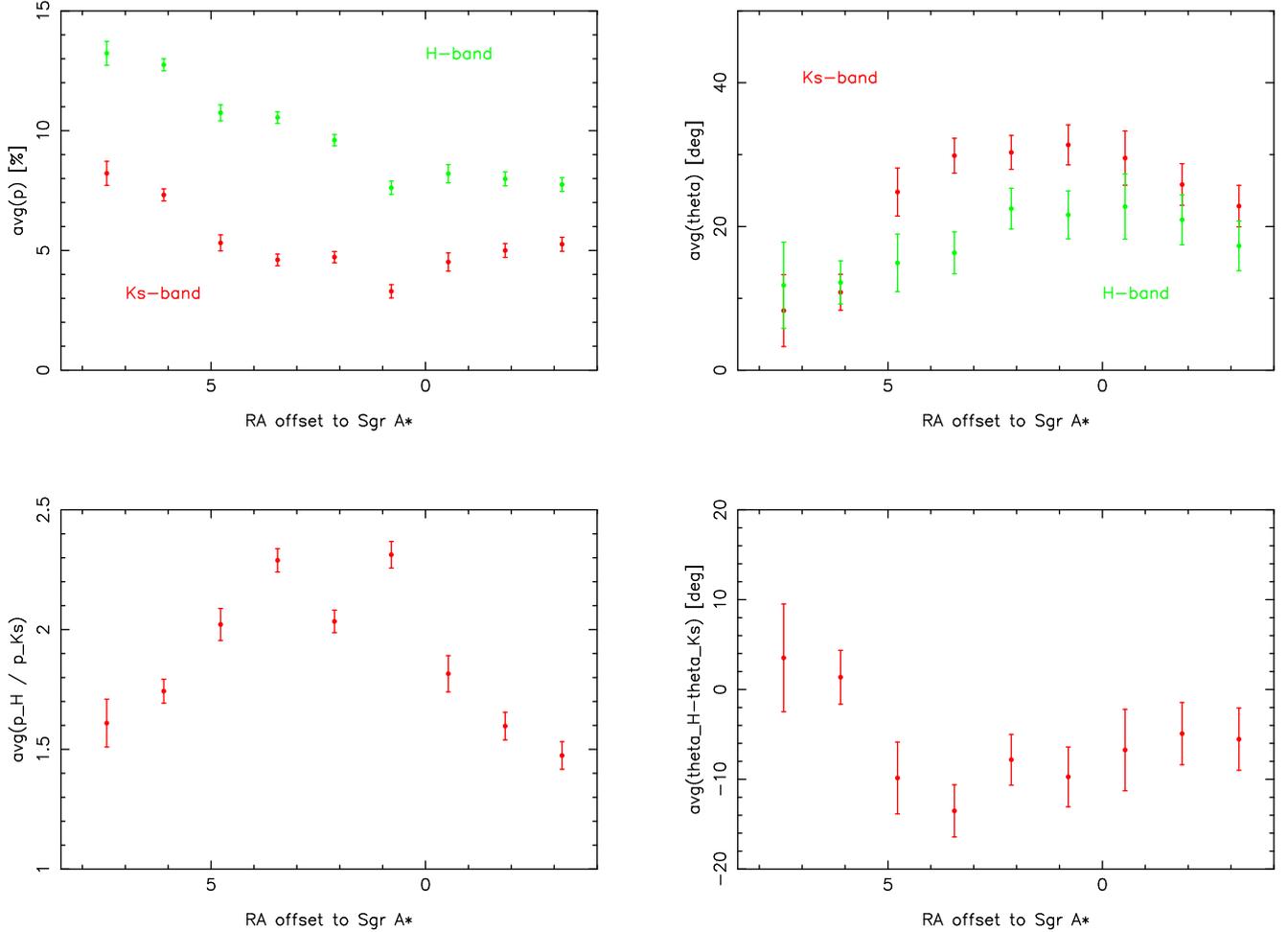}
\caption{\small H- and Ks-band polarization parameters, averaged over 1.3'' bins along the East-West axis. Upper left: polarization degrees, red: Ks-band, green: H-band. Upper right: polarization angles, red: Ks-band, green: H-band. Lower left: $\frac{p_H}{p_{Ks}}$.
Lower right: $\theta_H - \theta_{Ks}$.}
\label{FigPolparsbinned}
\end{figure*}

\begin{table}[!b]
\caption{\small Results of Gaussian fits to polarization parameter histograms for the complete dataset resp. sub-datasets separated
based on $p_{Ks}$ resp. position along the East-West-axis. Polarization degrees given in \%, angles given in degrees.}
\label{TabPolparsFull}
\centering       
\begin{tabular}{l l r r r r}
\hline\hline
separation & value & peak 1 & $\sigma_1$ & peak 2 & $\sigma_2$ \\
\hline
none & $p_{Ks}$[\%] & 4.6 & 0.8 & 7.7 & 1.2\\
none & $p_{H}$[\%] & 9.8 & 0.7 &  & \\
none & $\theta_{Ks} [^{\circ}]$ & 28 & 3 & 12 & 6\\
none & $\theta_{H} [^{\circ}]$ & 20 & 8 &  & \\
none & $\frac{p_H}{p_{Ks}}$ & 1.9 & 0.4 &  & \\
none & $\theta_{H}-\theta_{Ks} [^{\circ}]$ & 2 & 8 &  & \\
$p_{Ks}$ & $p_{Ks}$[\%] & 4.6 & 0.6 & 7.5 & 1.0\\
$p_{Ks}$ & $p_{H}$[\%] & 9.3 & 1.3 & 12.1 & 2.1\\
$p_{Ks}$ & $\theta_{Ks} [^{\circ}]$ & 28 & 6 & 11 & 6\\
$p_{Ks}$ & $\theta_{H} [^{\circ}]$ & 20 & 6 & 13 & 6\\
$p_{Ks}$ & $\frac{p_H}{p_{Ks}}$ & 2.0 & 0.3 & 1.6 & 0.3\\
$p_{Ks}$ & $\theta_{H}-\theta_{Ks} [^{\circ}]$ & -5 & 4 & 4 & 5\\
E-W pos & $p_{Ks}$[\%] & 4.6 & 0.6 & 7.7 & 1.0\\
E-W pos & $p_{H}$[\%] & 9.3 & 1.4 & 11.8 & 2.1\\
E-W pos & $\theta_{Ks} [^{\circ}]$ & 29 & 6 & 10 & 7\\
E-W pos & $\theta_{H} [^{\circ}]$ & 21 & 6 & 13 & 6\\
E-W pos & $\frac{p_H}{p_{Ks}}$ & 2.0 & 0.3 & 1.7 & 0.3\\
E-W pos & $\theta_{H}-\theta_{Ks} [^{\circ}]$ & -5 & 4 & 3 & 5\\
\hline
\end{tabular}
\end{table}
By comparing the positions of the sources detected in each datasets, we found 133 sources with reliable polarization parameters
common to the H- and the Ks-band data. The missing sources are mostly found outside the other data-set's FOV, in
addition to a small number of very fast moving sources (like the S stars), which are difficult to identify owing to the
time of 2.5 years between the H- and Ks-band observations. In addition, photometric errors tend to be larger in the westernmost 
region of the FOV because of the lack of suitable close bright stars for PSF determination. This leads to less sources with
reliable polarization parameters there. The common sources are used for the following source-by-source comparison.\\
We find that the polarization angles measured in the H- and Ks-band agree well in the eastern part of the FOV, while there appears
to be an offset in the center and the western region. Fig.\ref{FigPolcompHKtot} shows the
difference in polarization angle and the relation of H- and Ks-band polarization degrees. $\theta_{H}-\theta_{Ks}$ can be fitted
well with a Gaussian distribution and shows a peak at $4^{\circ} \pm 8^{\circ}$ (FWHM of 20$^{\circ}$). Considering the width of the
peak, this offset is not significantly different from zero. $\frac{p_H}{p_{Ks}}$ can be fitted quite well with a log-normal distribution,
peaking at 1.9 $\pm$ 0.4 (FWHM of 0.9). For a complete list of the fitting results, see Tab.\ref{TabPolparsFull} (which also contains
the values referred to in the paragraphs below).\\
Assuming the two peaks we find for both the polarization degree and angle in the Ks-band are real, we separated
the stars with reliable polarization parameters in both H- and Ks-band into two samples: stars with $p_{Ks}<6\%$
(pK$^-$) and with $p_{Ks}>6 \%$ (pK$^+$). Fig.\ref{FigPolcompHKpolsep} shows histograms of the different polarization
parameters of the pK$^-$ and pK$^+$ sources in the two bands, namely the polarization degrees and angles, $\theta_H-\theta_{Ks}$
and $\frac{p_H}{p_{Ks}}$. All histograms were fitted with a Gaussian, and although these fits are poor in several cases, the fitted
peaks show at least the trends present in the data.\\
We find that the pK$^+$ sources show systematically lower polarization angles than the pK$^-$ sources in the Ks-band (peak offset
of 18$^{\circ}$). The peaks fitted here match the ones fitted to the complete dataset (see \S \ref{SectKpol}). A similar, yet smaller
offset exists in the H-band (7$^{\circ}$, well within the uncertainties, the fitted peaks also correspond to those determined in
\S \ref{SectHpol}). Accordingly, $\theta_{H}-\theta_{Ks}$ also shows an offset of 9$^{\circ}$, but that value is relatively close
to zero for both sub-datasets considering the large FWHM of the peaks.\\
Looking at the polarization degrees, we find similar offsets: for the Ks-band, the pK$^+$ peak is found at a polarization degree
which is higher by a factor of 1.6 than where the pK$^-$ peak is fitted. The relative difference found in the H-band is smaller,
with the pK$^+$ peak found at a polarization degree which is higher by a factor of 1.3 than that of the pK$^-$ peak. This manifests
itself in the $\frac{p_H}{p_{Ks}}$ histogram, which we find peaked at 1.6 for pK$^+$, and at 2.0 for pK$^-$.\\
Because we mostly find the higher polarized pK$^+$ sources in the eastern part of our FOV (in the general area of the
Northern Arm), we also divided our sources into two samples based on their position along the East-West-axis: pK-E (sources
more than 4.1'' east of Sgr A*) and pK-W (sources less than 4.1'' east of Sgr A*). Fig.\ref{FigPolcompHKxpos} shows histograms
of the polarization parameters for both sub-datasets. We find results that are very similar to those found for a separation based
on $p_{Ks}$, with practically identical peaks and offsets (pK-E corresponding to pK$^+$, pK-W to pK$^-$).\\
How can we explain these findings? For interstellar polarization in general, the H-band polarization degrees are expected to be
significantly higher than the Ks-band values, while the angles in both bands should be the same within the uncertainties.
This is expected from the Serkowski law and the power law relation presented by \cite{martin1990}. According to the semi-empirical
Serkowski law, the polarization at a given wavelength in relation to the polarization maximum depends on the wavelength
where that maximum occurs
\begin{equation}
\frac{p(\lambda)}{p_{max}} = exp \left[ -K \times ln^2 \left( \frac{\lambda_{max}}{\lambda} \right) \right]
\end{equation}
with $K = 0.01 + 1.66 \lambda_{max}$ \citep{whittet1992}. It appears, however, that polarization in
the NIR, specifically in the J-, H- and K-band, only very weakly depends on $\lambda_{max}$ \citep{martin1990}.
Keeping this in mind and considering the availability of only two data-points for each source and the large FWHM of the Gaussian 
fits to $\frac{p_H}{p_{Ks}}$, we can only give rough estimates here. We find that the peak fitted to the complete dataset
(1.9 $\pm$ 0.9) agrees best with $\lambda_{max} \sim 0.7 \mu m$. Of the two fitted peaks for the sub-datasets, the value of
2.0 $\pm$ 0.7 would agree with $\lambda_{max} \sim 0.7 \mu m$ as well, while the peak at 1.6 $\pm$ 0.7 points to a $\lambda_{max}$
which is either much smaller ($\sim 0.25 \mu m$) or larger ($\sim 1.25 \mu m$). \cite{bailey1984} give a value of
$\lambda_{max} \approx 0.80 \mu m$ (using J, H-and K-band data), which approximately matches the first two values we find here.
The authors state that the agreement with the semi-empirical law is only rough. It also has to be considered that
this law was established for sources with only very weak polarization in the NIR, and consequently does not describe observations in that
wavelength regime very well. All this leads us to conclude that our data clearly are not sufficient to give a reliable estimate
for this parameter.\\
Using the power-law relation proposed by \cite{martin1990}
\begin{equation}
\frac{p_H}{p_{Ks}} = \left( \frac{\lambda_H}{\lambda_{Ks}} \right) ^{-\alpha}
\end{equation}
we obtain a power-law index of $\alpha = 2.4 \pm 1.7$ for $\frac{p_H}{p_{Ks}} = 1.9 \pm 0.9$, while the two sub-dataset peaks lead to
$\alpha = 1.7 \pm 1.6$ resp. $\alpha = 2.5 \pm 1.3$. These values agree with the range of 1.5-2.0 given by \cite{martin1990},
although the uncertainties are quite large. It has to be stressed that our values have been obtained on a relatively narrow region
and that a study of a much wider region is necessary before any reliable conclusions can be drawn on this matter.\\
Fig.\ref{FigPolparsbinned} shows binned plots of the H- and Ks-band polarization parameters. Only the common reliable sources were
used for this comparison. We averaged the polarization parameters for all sources contained in 1.3'' wide bins along the East-West
axis.\\
We observe the same large-scale trends found based on the histograms, and the first impression points toward the same effects
being present in both Ks- and H-band for both polarization degree and angle. Looking at the plots of $\frac{p_H}{p_{Ks}}$ and
$\theta_H-\theta_{Ks}$, however, the differences between the two bands become apparent: as seen in Fig.\ref{FigPolcompHKxpos}, 
$\frac{p_H}{p_{Ks}}$ is found to be around 2-2.2 in the center and more toward 1.5-1.6 east of Sgr A*. A similar trend is
visible toward the western edge, but there are few sources there and there may be an influence of edge effects.
The $\theta_H-\theta_{Ks}$ plot also confirms the trends found earlier: we see an offset of $\sim$10$^{\circ}$ in the center and the 
west and one of $\sim$0$^{\circ}$ toward the east.\\  
These results raise the question for the cause of the observed deviations of the polarization parameters over the FOV, assuming
those deviations are indeed real and not some sort of instrumental effect we did not account for. The comparison of our main Ks-band
dataset (2) with dataset 4 shows that while a similar increase in polarization degree exists there as well, no accompanying shift
in polarization angle is found. If this was indeed an instrumental effect, one would expect the same pattern in both parameters.
Fig.\ref{FigIRS16Ccomp} shows a residue of the dither pattern in both Q and U, but these variations are far too small to lead
to a significant deviation over the FOV as we observe it here.\\ 
It has to be considered that the significance of the trends we find stays below $3\sigma$, based on the systematic
uncertainties determined in \S \ref{SectKpol}\\
This question cannot be solved completely without additional (calibration) observations, but for the purposes of this study, we assume
that what we observe is indeed a real effect.
As possible explanations, two basic mechanisms come to mind here:
\paragraph{Variable LOS extinction}
The extinction toward the central parsec is known to be ''patchy'' \citep[see][]{schoedel2010b}, which in turn indicates
different dust column densities and/or dust parameters along individual lines-of-sight toward different regions in the FOV. This
could lead to differences in the polarization measured at different locations. But the situation is even more complex. Not only
the densities are important, but also possible different alignment in individual dust clouds (so passing through an additional
cloud on one LOS compared to another could even lower the total polarization degree for that LOS). Another problem is that we
find a smaller effect in the H-band (and thus the impact on $\theta_H-\theta_{Ks}$ and $\frac{p_H}{p_{Ks}}$). This would
require significantly different average dust parameters from one LOS
to the other (which in turn requires even more dramatic
changes for a considerable percentage of the individual dust clouds along the LOS). Such a configuration is possible, but the highly
specific arrangement required to produce a pattern as we observe it seems very unlikely.
\paragraph{Local influences}
The area where this effect occurs coincides with the position of a known local feature, the Northern Arm of the
Minispiral. This feature is clearly visible in the L-band (see Fig.\ref{FigLband}), faint in Ks and not
detectable in the H-band. The light from the stellar sources itself, which passes through the stream of aligned grains in
the Northern Arm, and scattered and/or emitted light from these grains themselves may contribute to some extent to the
polarization measured in this region. This would have a stronger impact in the Ks-band compared to the H-band, because of the
sizes and temperatures of the involved grains, this far matching our findings. The question however remains how substantial such a
contribution could be.\\
Under conditions as they are found in the filaments, grain alignment by the Davis-Greenstein mechanism would be almost perfect,
especially because of the strong magnetic fields \citep[lower limit of $\sim$2 mG close to IRS~1W, see][]{aitken1998}. By coincidence,
the local grain alignment angle in the Northern Arm matches that measured for the LOS polarization \citep{aitken1998}.
This means that local dichroic extinction would increase the observed total polarization degree, while scattering/emission would
decrease it (as is indeed found for IRS~1W, see below).\\
While the trends in polarization degree in the 2009 data may be explained this way, the measured polarization angles do not show
the expected behavior: if local and LOS dichroic extinction produce alignment in the same direction, the polarization angles
should follow the magnetic field lines (see Fig.\ref{FigPolmapK}). This is the case between 5'' and 1'' west of Sgr A*, but
farther west, the angles are $\sim 20^{\circ}$ steeper than what would be expected. In the H-band, the angles are less steep in that
region, providing a better match to the field lines. The preliminary Ks-band results (March 2011) also seem to agree well
with the field. Farther to the south, however, the higher polarization degrees in the 2007 Ks-band data-set occur despite a
complicated field structure that does not offer an easy explanation for this effect.\\
It is questionable whether local dichroic extinction can account for these effects, or if other contributions from scattering or
extended emission play a role here. The region observed here is quite complex, with a complicated field structure and gas/dust
streams, and in addition, the exact position of the observed sources along the LOS axis is not clear. It is therefore difficult
to determine what effect even a given dust distribution with a known temperature and composition would have on a certain source,
but not even these parameters are known exactly.\\ 
We therefore conclude that local effects, such as local dichroic extinction, extended emission by and scattering on dust grains 
are the most likely cause for this effect, although we cannot determine the likely extent of the individual contribution from
the available data on dust parameters and configurations along the LOS and in the GC itself.\\
This immediately leads to the next question: are these trends mirrored by the behavior of the extinction measured in the FOV? 
\begin{figure*}[!t]
\centering
\includegraphics[width=\textwidth,angle=0, scale=1.0]{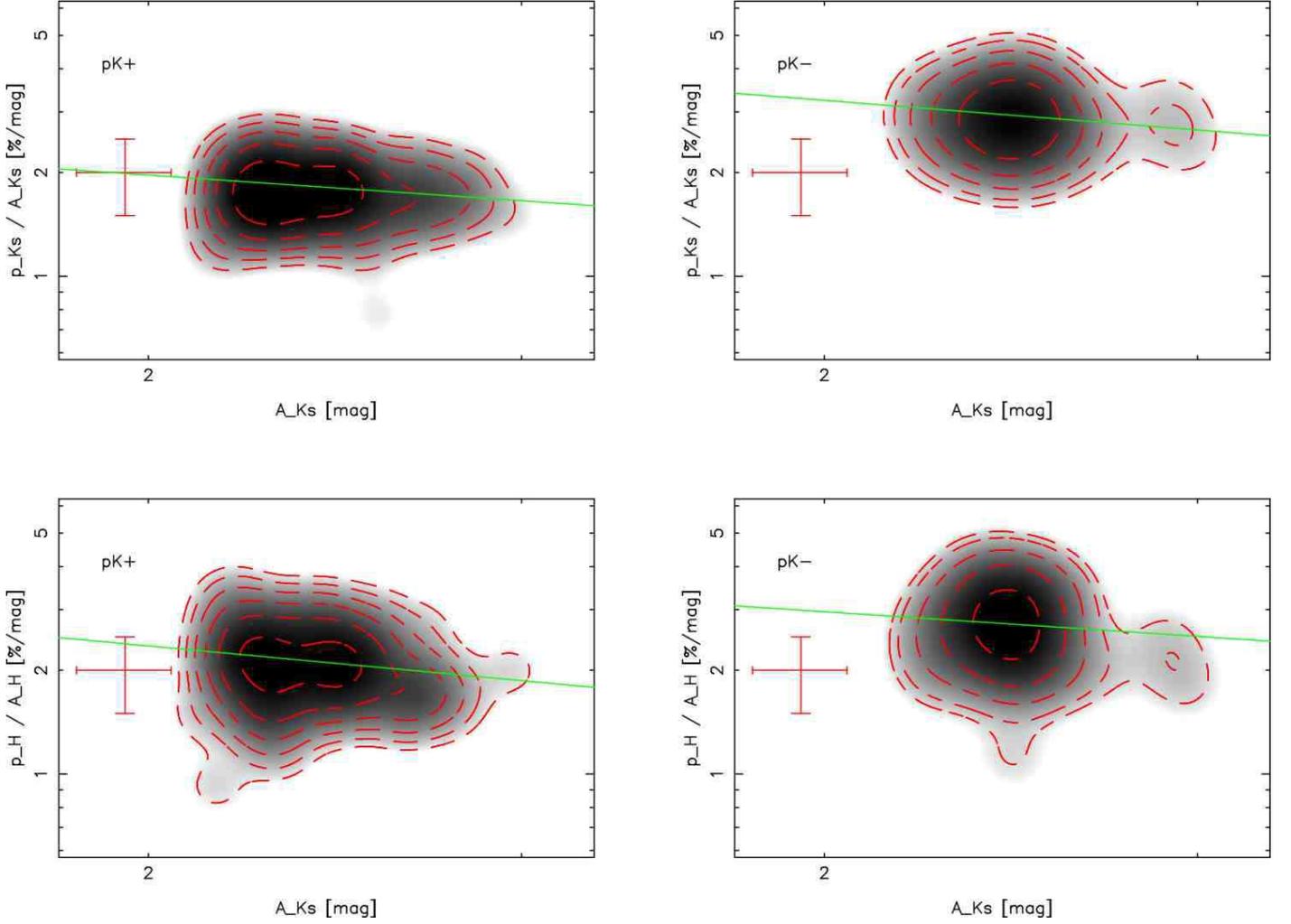}
\caption{\small Polarization efficiency in Ks-band (upper frames) and H-band (lower frames), compared to Ks-band extinction,
plotted as point density, with typical error represented in upper left corner. pK$^+$ and pK$^-$ sources shown separately in
left resp. right frames for both bands. Green lines represent the fitted power-law relation.}
\label{FigPoleff}
\end{figure*}
\subsection{Correlation with extinction}
\label{SectExtinction}
\begin{figure}[!b]
\centering
\includegraphics[width=\textwidth,angle=-90, scale=0.30]{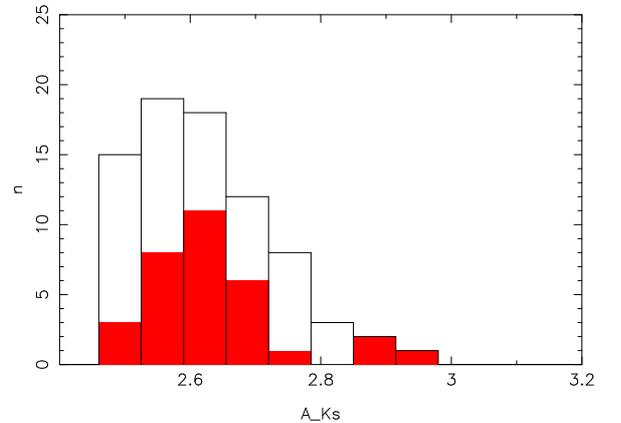}
\caption{\small Ks-band extinction for sources with p$_{Ks} < 6$\% (open columns) resp. p$_{Ks} > 6$\% (filled columns).}
\label{FigExtpolsep}
\end{figure}

We compared the H- and Ks-band polarization values to the extinction map presented by \cite{schoedel2010b}. Fig.\ref{FigPoleff}
shows the polarization efficiency $\frac{p_{\lambda}}{A_{\lambda}}$ for both bands plotted against the Ks-band extinction $A_{Ks}$
taken from the extinction map at the location of each source. As it turns out, almost the same distribution of $A_{Ks}$ is found
for the pK$^+$ and the pK$^-$ sources (see Fig.\ref{FigExtpolsep}), which in turn leads to an offset between the two sub-datasets
in polarization efficiency. We therefore plot them separately in Fig.\ref{FigPoleff}.
In both bands, the distributions can be fitted
with a power law:
\begin{equation}
\frac{p_{\lambda}}{A_{\lambda}} \propto A_{Ks} ^{\beta},
\end{equation}
with $\beta_{Ks,-} = -0.4 \pm 0.4$ resp. $\beta_{H,-} = -0.6 \pm 0.6$ in the H-band for the pK$^-$ sources. Despite the
large errors which stem from considerable scatter of the parameters, this matches the relation found by \cite{gerakines1995}
for the Taurus dark cloud and also the results of \cite{whittet2008} for Ophiuchus (both in the Ks-band). This is not directly
comparable, because the GC is obscured by more than one dust cloud with possibly different dust alignments. A more general study
by \cite{jones1989}, examining a great number of sources covering a range in optical depth of about a factor of 100, finds a
power law relation with $\beta \sim -0.25$. In this study, the author proposed a model where the magnetic field along the LOS consists
of a constant and a random component \citep[see also][]{heilis1987}, thus leading to different grain alignment in each section
along the LOS. This reproduces the findings in that study quite well, and it is also consistent with our own results within the
uncertainties.\\
For the smaller number of pK$^+$ sources, we find similar power law indices: $\beta_{Ks,+} = -0.5 \pm 0.7$
resp. $\beta_{H,+} = -0.4 \pm 1.5$. Compared to the pH/K$^-$ values, we find a significant offset in polarization efficiency,
while the underlying power law appears to be very similar.
This might indicate that the additional polarization is indeed
caused by a local contribution, likely of Northern Arm material. To produce this deviation along the LOS, a very specific
and therefore unlikely dust configuration would be required.  
\subsection{Examining the extended sources}
\label{SectExtended}
Our polarimetric data cover two bright extended sources in the central parsec, IRS~1W and IRS~21. The former is contained
in both our H- and Ks-band data (datasets 1 and 2), while the latter is covered by several rotated Wollaston datasets of poorer
quality (datasets 3-15). IRS~1W shows a clear horseshoe shape as expected for a bow-shock source in high-quality Ks-band
images, while IRS~21 does not. Owing to its high apparent polarization, IRS~21 would be interesting as a polarimetric calibration
source, but only if the polarization is not variable. To constrain the nature of this source even more, we also conducted
a flux variability analysis on IRS~21 and other extended sources in the FOV of our polarimetric data and several NACO imaging
datasets, taken between 2002 and 2009 in the H-, Ks- and L-band.
\begin{figure}[!t]
\centering
\includegraphics[width=\textwidth,angle=-90, scale=0.4]{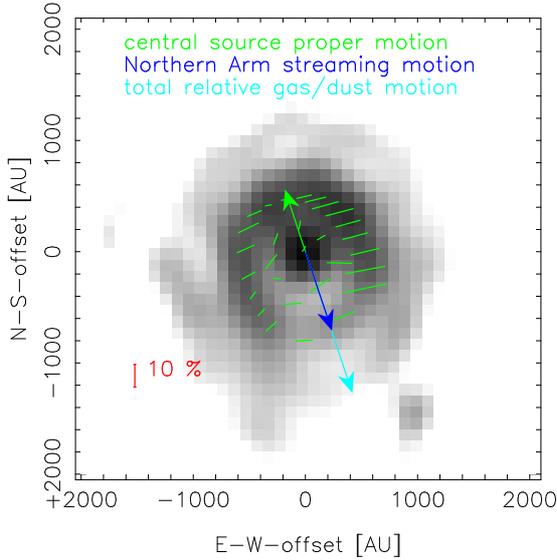}
\caption{\small Map of the intrinsic Ks-band polarization of the extended source IRS~1W. The arrows indicate the proper motions
of the central source, the motion of the Northern Arm material and the motion of both relative to each other.}
\label{FigIRS1W2dKs}
\end{figure}
\subsubsection{IRS~1W}
\paragraph{Source morphology}
IRS~1W shows the characteristic horseshoe-shape of a bow-shock source \citep[see e.g.][]{tanner2005}. This shape can already be
made out in the raw images, but it becomes even more apparent in a Lucy-Richardson deconvolved image (using a PSF obtained from
bright IRS~16 sources, see Fig.\ref{FigIRS1W2dKs}). The observed shape agrees very well with
the relative velocities of the streaming material of the Northern Arm \citep{lacy1991} at the location of the source and the
proper motion of IRS~1W itself \citep[][, velocities plotted in Fig.\ref{FigIRS1W2dKs}]{schoedel2009}. 
\begin{figure}[!t]
\centering
\includegraphics[width=\textwidth,angle=-90, scale=0.4]{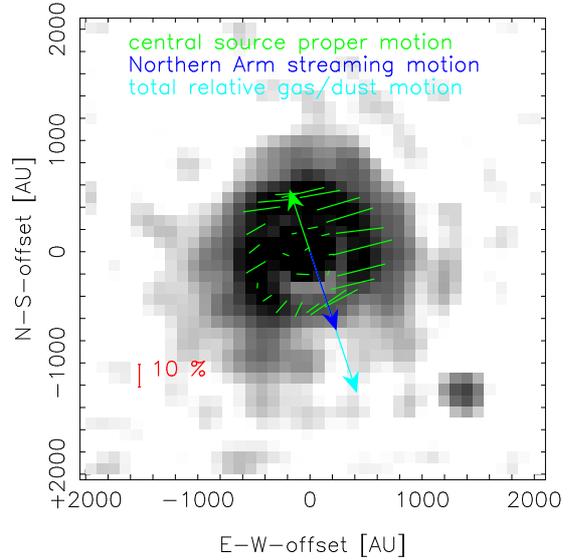}
\caption{\small Map of the intrinsic H-band polarization of the extended source IRS~1W (see Fig.\ref{FigIRS1W2dKs}).}
\label{FigIRS1W2dH}
\end{figure}
\paragraph{Spatially resolved polarimetry}
We measured the total polarization of IRS~1W as (1.8 $\pm$ 0.5) \% at (-37 $\pm$ 5)$^{\circ}$ East-of-North after
application of the M\"uller matrix to account for instrumental polarization. \cite{ott1999} provide values of (4.6 $\pm$
2.5) \% at (-85 $\pm$ 8)$^{\circ}$ East-of-North. It has to be considered that instrumental effects take place on the same
order of magnitude as the measured polarization degree, which may explain the large offset in polarization angle compared
to the older values where these effects were not compensated. We attribute our lower total polarization degree to the
better angular resolution and thus less influence from neighboring sources. The values provided by \cite{eckart1995} are
clearly influenced by neighboring stellar sources, with (3.0 $\pm$ 1.0) \% at (10 $\pm$ 7)$^{\circ}$.\\
\begin{figure}[!b]
\centering
\includegraphics[width=\textwidth,angle=-90, scale=0.35]{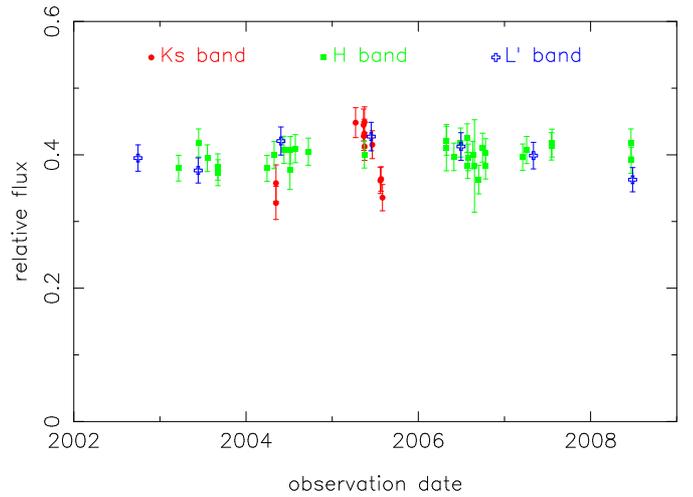}
\caption{\small Ks-band (red circles), H-band (green squares) and L-band (blue open crosses) lightcurves of IRS~1W. Shown
here is the flux of the source normalized to the flux of non-variable reference sources in the respective band. H- and
L-band values were multiplied by a constant factor to match the average Ks-band values.}
\label{FigIRS1Wfluxvar}
\end{figure}
If we assume that the foreground polarization
for this source is the same as for the surrounding objects and apply a depolarization matrix with the parameters for $p =
7.6 \%, \theta = 9.2^{\circ}$, we find a total intrinsic polarization of (7.8 $\pm$ 0.5) \% at (-75 $\pm$ 5)$^{\circ}$.\\
In the H-band we measured the total polarization of IRS~1W as (5.2 $\pm$ 0.5) \% at (12 $\pm$ 5)$^{\circ}$ East-of-North.
The polarization angle appears typical for a stellar source affected by foreground polarization, but the polarization
degree is much lower than the $\sim$12\% found for stellar sources in the vicinity. We applied a depolarization matrix
with  $p = 12 \%, \theta = 15^{\circ}$ and found a resulting intrinsic polarization of (6.9 $\pm$ 0.5) \% at (-73 $\pm$ 5)$^{\circ}$.
The angle agrees very well with the Ks-band polarization angle, suggesting that the same process is responsible.
The lower intrinsic polarization degree points to the lower influence of the extended dust component compared to that of the
central source at this wavelength.\\
\begin{figure*}[!t]
\centering
\includegraphics[width=\textwidth,angle=-90, scale=0.40]{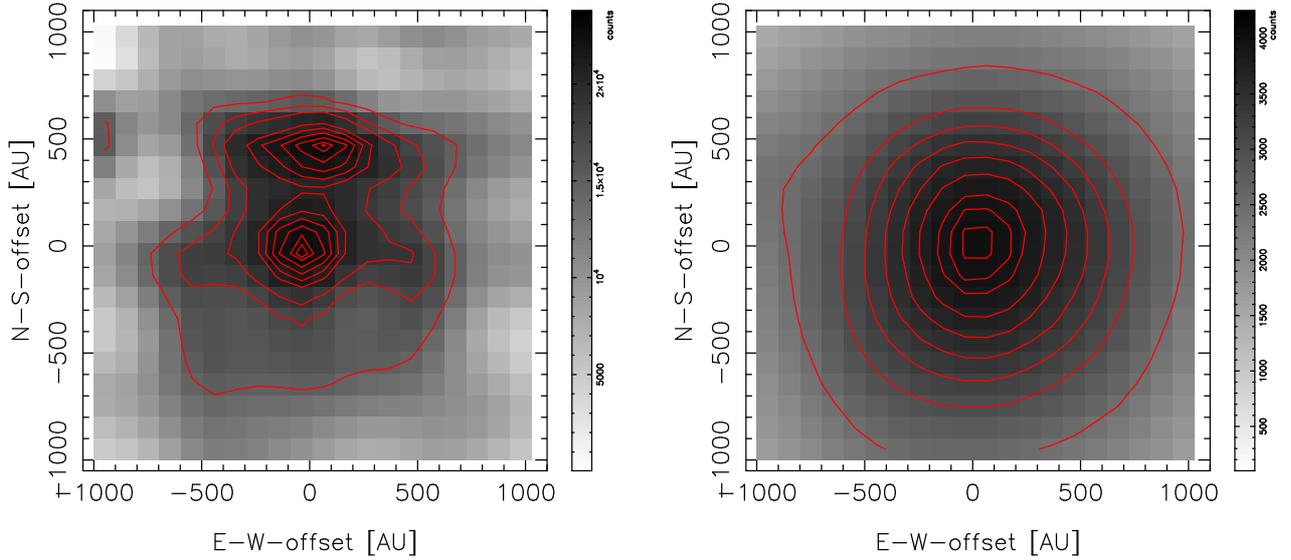}
\caption{\small IRS~21 observed on 2004-08-30, right: image before Lucy-Richardson deconvolution, left: image after
deconvolution (logarithmic gray-scale, contours trace 10, 20, 30, 50, 60, 70, 80, 90, 95\% of maximum flux in the left
frame, resp. 40, 50, 60, 70, 80, 90, 95, 99\% in the right frame.}
\label{FigIRS21resolved1}
\end{figure*}
\begin{figure*}[!t]
\centering
\includegraphics[width=\textwidth,angle=-90, scale=0.40]{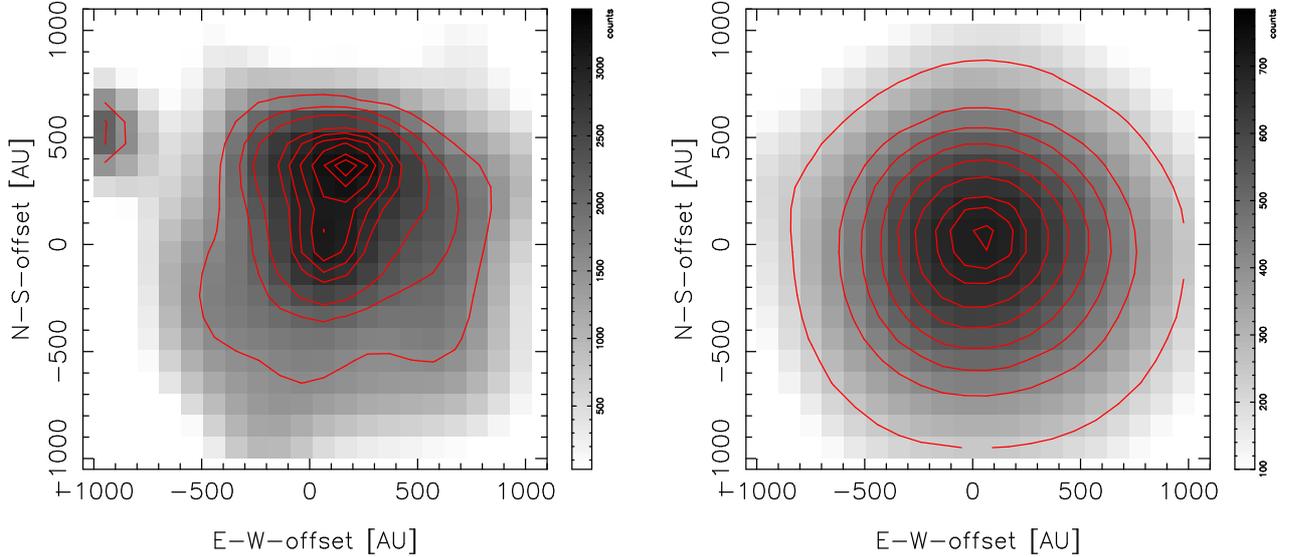}
\caption{\small IRS~21 observed on 2005-05-14, right: image before Lucy-Richardson deconvolution, left: image after
deconvolution (logarithmic gray-scale, contours trace 10, 20, 30, 50, 60, 70, 80, 90, 95\% of maximum flux in the left
frame, resp. 40, 50, 60, 70, 80, 90, 95, 99\% in the right frame.}
\label{FigIRS21resolved2}
\end{figure*}
We measured the polarization of individual regions of the source in the deconvolved images and subtracted the foreground polarization
using a M\"uller matrix as described in \S \ref{SectForegroundremoval}. The results are shown in Fig.\ref{FigIRS1W2dKs}. We find
polarization degrees of about 10-20 \% with very similar polarization angles for regions with significant flux. Apparently,
the polarization degree is lower by a factor of up to $\sim$3 around the apex compared to the tails.\\
We applied the same technique to our H-band dataset, and Fig.\ref{FigIRS1W2dH} shows the result for the immediate area
around IRS~1W. We find a polarization pattern comparable to that in the Ks-band, with polarization degrees of about 10-20
\% with very similar polarization angles for regions with significant flux and much lower polarization degrees in the
central region. There is no significantly lower polarization degree toward the apex, as found in the Ks-band, but
it has to be noted that the horseshoe shape is much less pronounced in the H-band anyway. The angles agree with those determined
from the Ks-band data.\\
The total intrinsic polarization and the spatially resolved pattern of IRS~1W can be explained from the combination of the
motion of the source itself, the streaming velocity of the Northern Arm and the magnetic fields present in that structure.\\
\cite{aitken1998} mapped the polarization of the Northern Arm at 12.5 $\mu$m and inferred the magnetic field orientation by assuming
that the polarization was produced by emission from magnetically aligned elongated dust grains. The magnetic field at the location
of IRS~1W is perpendicular to the polarization angles we find here. The projected velocity of the source \citep{schoedel2009}
is parallel to the field lines and also parallel to the streaming motion in the Northern Arm itself \citep{lacy1991}. This leads to
the field lines following the morphology of the shock around the source. At the apex this causes a weakening of the field, while it
is compressed in the tails. This in turn leads to a weaker resp. stronger grain alignment in the apex resp. the tails.\\
Grain temperatures in the Northern Arm reach up to $\sim$200-300 K \citep{smith1990,gezari1992}, and this is sufficient to explain the 
observed 12.5 $\mu$m emission.
But for significant emission in the H- and Ks-band, much higher grain temperatures of $\sim$1000 K would be required. This raises
the question if the extended emission in the bow-shock is emission from or scattering on the aligned grains. One possibility
is that the temperature in the shock is indeed high enough for significant emission. \cite{geballe2004} suggest that these higher
temperatures could be reached by very small grains ($\sim$0.001-0.01 $\mu$m, which is by a factor of 10 smaller than typical grains
expected here) if they are heated by occasional high-energy photons or stochastic collisions with high-energy electrons or ions.
\cite{moultaka2004} find that the spectrum measured for IRS~1W (including the bow-shock) matches a 900 K blackbody. At this higher
temperature, emission from heated aligned dust grains should contribute to the Ks-band and H-band polarization.\\
\begin{figure}[!t]
\centering
\includegraphics[width=\textwidth,angle=-90, scale=0.4]{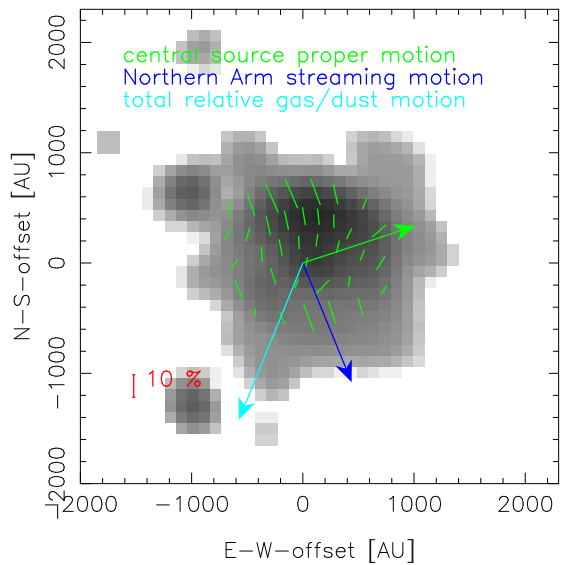}
\caption{\small Map of the intrinsic Ks-band polarization of the extended source IRS~21 (see Fig.\ref{FigIRS1W2dKs}).}
\label{FigIRS21_2d}
\end{figure}
This leaves the question of the survival of these very small grains in a bow-shock environment and at this temperature. In addition,
these grains would have to be aligned to produce the observed polarization. But the very same processes
that would heat the small grains to high temperatures would also randomize any previous alignment unless the alignment mechanism
is much faster than the randomization. This may be the case here, because strong frozen-in and compressed magnetic fields
provide an even stronger and faster alignment than in other regions of the Northern Arm. This might also add
to the lower polarization at the apex due to turbulence, which could lead to a partial randomization of grain alignment, before the
stronger field and uniform streaming motion in the tails increase the alignment again.\\
In addition to emission, scattered light from the central source enclosed in a dusty envelope could contribute to the observed
polarization. This process is known to occur in planetary nebulae and dusty young stellar objects (YSOs) \citep[e.g.][]{lowe2007,lucas1998}.
\begin{figure}[!t]
\centering
\includegraphics[width=\textwidth,angle=-90, scale=0.35]{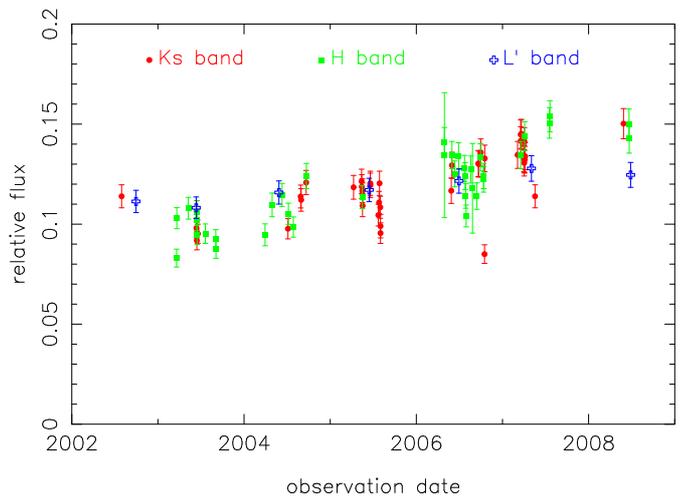}
\caption{\small Ks-band (red circles), H-band (green squares) and L-band (blue open crosses) lightcurves of IRS~21. Shown
here is the flux of the source normalized to the flux of non-variable reference sources in the respective band. H- and
L-band values were multiplied by a constant factor to match the average Ks-band values.}
\label{FigIRS21fluxvar}
\end{figure}
\begin{table}[!b]
\caption{\small Polarization parameters measured for IRS~21. The average over all observations resp. the standard
deviation is given in the last line.}
\label{TabIRS21pol}
\centering       
\begin{tabular}{l l r r r r}
\hline\hline
 & date & p$[\%]$ & dp$[\%]$ & $\theta$ [deg] & d $\theta$ [\%] \\
\hline
1 & 2007-04-01 & 10.4 & 0.5 & 15 & 5\\
2 & 2007-04-03 & 9.3 & 0.5 & 16 & 5\\
3 & 2007-04-04 & 8.9 & 0.5 & 15 & 5\\
4 & 2007-04-05 & 8.8 & 0.5 & 17 & 5\\
5 & 2007-04-06 & 9.0 & 0.5 & 16 & 5\\
6 & 2007-07-18 & 9.4 & 0.5 & 16 & 5\\
7 & 2007-07-19 & 9.0 & 0.5 & 17 & 5\\
8 & 2007-07-20 & 9.1 & 0.5 & 16 & 5\\
9 & 2007-07-20 & 9.0 & 0.5 & 18 & 5\\
10 & 2007-07-21 & 8.4 & 0.5 & 17 & 5\\
11 & 2007-07-23 & 9.2 & 0.5 & 16 & 5\\
12 & 2007-07-23 & 8.3 & 0.5 & 18 & 5\\
13 & 2007-07-24 & 9.5 & 0.5 & 16 & 5\\
avg & & 9.1 & 0.2 & 16.4 & 0.3\\
\hline
\end{tabular}
\end{table}
In the ideal case of pure Mie-scattering on spherical dust grains, most of the light would be scattered forward, but a significant
portion is scattered perpendicular to the incident direction. This latter part is linearly polarized, with the polarization vector
in the plane of the sky and perpendicular to the original propagation direction of the light. If a source like this is viewed
face-on, no total intrinsic polarization is detected, because the polarizations of the regions surrounding the central source
cancel each other out.
If the source is inclined, any total polarization angle can be produced. Spatially resolved
polarimetric measurements of these sources show a characteristic centrosymmetric pattern of the polarization vectors, however.
Clearly, this is not the kind of pattern we find here.\\
But what if the grains are not spherical, but elongated and aligned as is the case here at least for a significant part of the
grain population? \cite{lucas1998} find patterns of aligned polarization vectors
similar to the ones we find here in the central regions of a minority of the sources examined in their study. They claim that
these patterns cannot be explained by scattering on spherical grains, but that aligned elongated grains must play a role
there. \cite{whitney2002} modeled scattering and dichroic extinction for non-spherical dust grains, finding a variety of polarization
patterns for different input values for optical depth, degree of grain alignment, and inclination of the source. In general,
at low optical depths they find polarization vectors perpendicular to the axis of grain alignment (i.e. the orientation of the angular
momentum vector of the spinning grains), while high optical depth
leads to a predominance of dichroic extinction and thus to polarization vectors parallel to the axis of grain alignment.
This study is focussed on spherical dust configurations and disk-like structures, so the results are not directly applicable
to a bow-shock source.\\
\begin{figure}[!t]
\centering
\includegraphics[width=\textwidth,angle=-90, scale=0.35]{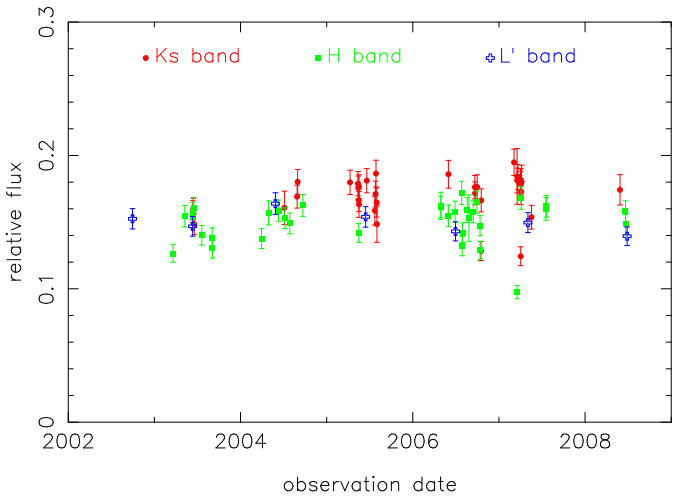}
\caption{\small Ks-band (red circles), H-band (green squares) and L-band (blue open crosses) lightcurves of IRS~10W. Shown
here is the flux of the source normalized to the flux of non-variable reference sources in the respective band. H- and
L-band values were multiplied by a constant factor to match the average Ks-band values.}
\label{FigIRS10Wfluxvar}
\end{figure}
In the light of these results, we consider it likely that scattering on elongated grains contributes to the observed patterns in
the H- and Ks-band data. Both scattering and emission should produce polarization at the same angle. Without thoroughly modeling
the conditions in such a bow-shock environment, we cannot conclude which process is dominant, only that both probably contribute.
\begin{figure}[!b]
\centering
\includegraphics[width=\textwidth,angle=-90, scale=0.35]{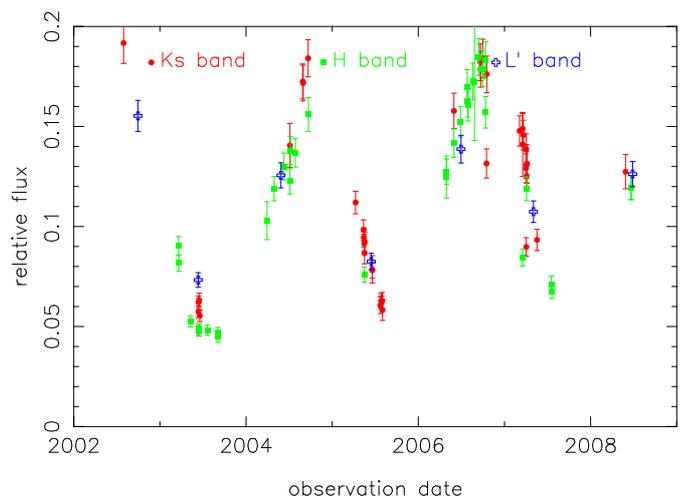}
\caption{\small Ks-band (red circles), H-band (green squares) and L-band (blue open crosses) lightcurves of IRS~10E. Shown
here is the flux of the source normalized to the flux of non-variable reference sources in the respective band. H- and
L-band values were multiplied by a constant factor to match the average Ks-band values.}
\label{FigIRS10Efluxvar}
\end{figure}
\paragraph{Flux variability}
The total flux of IRS~1W can be difficult to determine, because the center of this source is saturated in many images.
Unfortunately, the repairing algorithm included in StarFinder only works on point sources, because it assumes that
the saturated source has a PSF similar to that of non-saturated sources in the vicinity. This is obviously not the case
for an extended source like IRS~1W. We therefore only used data where the peak fluxes at the location of IRS~1W stayed below
the saturation threshold. Fig.\ref{FigIRS1Wfluxvar} shows the flux of IRS~1W in relation to the reference flux in the H-,
Ks- and L-band. The flux appears to be variable by $\sim$30\% in the Ks-band, while the H-band value is inconclusive.
No clear periodicity is apparent, which might indicate either an erratic variability or a period on the order of or below
the time resolution of our data. The flux variability exceeds that found for IRS~16C (see Fig.\ref{FigNonvar}), but this may in part be
due to the difficult photometry on IRS~1W. In the L-band, the flux rises to a maximum in 2005 and drops again by $\sim$15 \%.
We consider this consistent with a constant flux within the uncertainties, especially since there is no correlated behavior over the
H-, Ks- and L-band. In total, we cannot state whether or not IRS~1W shows a source-intrinsic flux variability based on the available observations.
\begin{figure}[!t]
\centering
\includegraphics[width=\textwidth,angle=-90, scale=0.35]{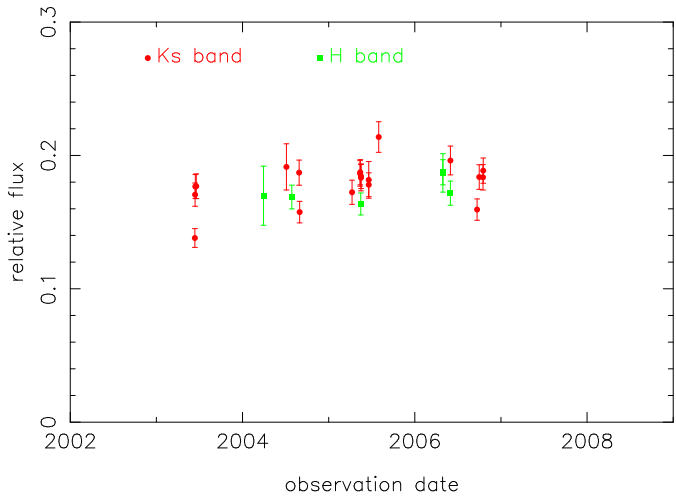}
\caption{\small Ks-band (red circles), H-band (green squares) and L-band (blue open crosses) lightcurves of IRS~5. Shown
here is the flux of the source normalized to the flux of non-variable reference sources in the respective band. H-band values were
multiplied by a constant factor to match the average Ks-band values.}
\label{FigIRS5fluxvar}
\end{figure}
\subsubsection{IRS~21}
\paragraph{Source morphology}
Previous studies show IRS~21 as a roughly circular, yet extended source \citep{tanner2002,tanner2005}. We find the same
for our data, but after applying a Lucy-Richardson deconvolution, this changes significantly: Fig.\ref{FigIRS21resolved1}
and \ref{FigIRS21resolved2} show IRS~21 before and after deconvolution for two datasets (2004-08-30 resp. 2005-05-14). There is
no clear bow-shock shape visible prior to deconvolution. After the LR process, there is still no clear horseshoe shape, but
it is clear that the source is not circular in projection and thus most likely not spherical. The deconvolved images
appear to contain a central source with a bow-shaped northern extension. It has to be noted that this shape is not this
apparent in all our datasets, but the resolution that can be achieved on an extended feature like this critically depends
on the data quality, especially in a very dusty environment like that of IRS~21. The LR algorithm also tends to 'suck up'
flux of extended features into a central source (see Appendix \ref{SectSimLR}). For all but a few periods, however, the
deconvolved images show an extended structure at this location, with an east-western bar/bow to the north and a
point-source-like feature to the south. We deem this shape to be consistent with a marginally resolved bow-shock type
source. Our findings match those presented by \cite{tanner2004}, who find a similar spatial structure and position angle of
the bow-shock in the L-band. Furthermore, the relative motion of the Northern Arm material \cite{lacy1991} and the proper
motions of IRS~21 \cite[e.g.][]{tanner2005} would agree with a bow-shock in this direction (see projected velocities plotted
in Fig.\ref{FigIRS21_2d}).\\
From the images where the bow-shock shape was visible, we determined a projected distance between the peaks of the southern
and the northern feature of $\sim 400 \pm 100$ AU (see Fig.\ref{FigIRS21resolved1}). The projected distance was almost the
same for all images, and the error
given here is just a rough estimate. Considering a possible inclination of the source, this constitutes a lower limit for the
standoff distance between the central source and the apex of the bow-shock. \cite{tanner2002} provide modeled standoff
distances for candidate central objects, using the Ks-band radius they measure for IRS~21 of $\sim$650 AU as an estimate for
the standoff distance. Using the radius makes this value an upper limit for the projected standoff distance. They find that
this value agrees best with the following stellar types: an AGB type star, W-R, Ofpe/WN9 or W-R WC9. Especially a
Wolf-Rayet (W-R) type star would be able to produce a wide range of possible standoff distances (310-10100 AU). Our value
agrees with these possibilities as well. 
\paragraph{Spatially resolved polarimetry}
IRS~21 is not covered by our main high-quality polarimetric dataset, so we measured its total polarization in several other
datasets with lower Strehl ratios (see Tab.\ref{TabObservations}, datasets 3-15, the data quality is still sufficient for
polarimetry on bright sources).
The values we find for the
individual datasets are shown in Tab.\ref{TabIRS21pol}. Considering the uncertainties, we conclude from these measurement
that both the polarization degree and angle of IRS~21 can be treated as constant within our margin of error. We find an
average value of (9.1 $\pm$ 0.2) \% at (16.4 $\pm$ 0.3) degrees, which agrees very well with the polarization
measured by \cite{ott1999} and also with the polarization angle determined by \cite{eckart1995}. The latter study measured a
polarization degree which is $\sim$60\% higher than our value. \cite{ott1999} explained this difference by different
apertures used and that might indeed be the case. We did not find a clear indication of a variable polarization degree of
IRS~21 in the available data, although several, low quality datasets led to some outlier values. Since this coincided with
large variations in the polarization of other sources, we do not consider this a significant effect.\\
After applying a depolarization matrix with polarization parameters determined from the surrounding point sources (5\% at 30$^{\circ}$,
determined on two sources close to IRS~21 and thus not as reliable as the values for IRS~1W),
IRS~21 appears to have a total intrinsic polarization of (6.1 $\pm$ 0.5) \% at (5 $\pm$ 5) degrees. The polarization
degree is slightly lower than that of IRS~1W, while both sources have very different intrinsic polarization angles. Again, the
angle we find is perpendicular to the magnetic field orientation in this region given by \cite{aitken1998}. Unfortunately
no polarimetric H-band data covering IRS~21 is available.\\
Fig.\ref{FigIRS21_2d} shows the polarization of individual regions. We find polarization degrees of about 3-8 \%, with less
uniform polarization angles than those found for IRS~1W in regions with significant flux. We also detect an increase instead of
a decrease of the polarization degree toward the apex. The polarization angle shows more variation compared to IRS~1W.\\
Again, the relation of the source motion and the local magnetic fields and streaming motions offer an explanation. The source
moves almost perpendicular to the field, so the field lines are likely compressed in front of the shock and diluted toward the
flanks. This leads to a higher polarization at the apex. If this field orientation is indeed preserved over the whole structure,
it could explain the observed polarization pattern, because the polarization angles would be perpendicular to the field lines. This would
be the expected behavior for emission/scattering producing the polarization.\\  
For this grain alignment dichroic extinction would be expected to produce polarization angles perpendicular to what is observed
here. This would reduce the overall intrinsic polarization, but it is questionable if the optical depth of the dust surrounding
IRS~21 is sufficient for a significant contribution of this process. Again, a detailed model of bow-shock polarization using
measurements at different NIR wavelengths as input parameters might clarify the extent of the influence of these processes. For
the purposes of this study, we assume that scattering and emission play the dominant role.
\paragraph{Flux variability}
Fig.\ref{FigIRS21fluxvar} shows the flux of IRS~21 in relation to the reference flux. For the H- and the Ks-band
data the flux significantly and steadily increases by about 50\% over the observed period (June 2002 to
May 2008). This corresponds to an increase in brightness of about 0.4 mag. That we find the same increase for
both bands suggests that it is intrinsic to the source itself, because the H-band is dominated by the stellar component,
while the Ks-band source is dominated by the dust enshrouding the stellar source. We find a lower ($\sim$20\%) correlated
flux increase in the L-band data as well. What could cause this increase in flux?\\
From the available data we cannot determine whether this is a periodic variability or not because we only observe a steady
increase in flux. This would be possible for all candidates presented by \cite{tanner2002}: both AGB-stars and Wolf
Rayet stars may show such an increase in luminosity in this time-frame, especially if they are in a mass-losing phase.\\
\begin{table*}[!t]
\caption{\small Details of the observations used for this work. N is
  the number of exposures that were taken with a given detector
  integration time (DIT). NDIT denotes the number of integrations that
  were averaged online by the read-out electronics during the
  observation. The Strehl ratio was measured using the {\em Strehl} algorithm
  of the ESO eclipse package (see N. Devillard, ''The eclipse software'',
  The messenger No 87 - March 1997, publicly available at {\em
  http://www.eso.org/projects/aot/eclipse/distrib/index.html}, given here is
  the average value over all images in each dataset.}
\label{TabObservations}
\centering       
\begin{tabular}{l l l l l l l l l l l}
\hline\hline
 & program & date & band & mode & channels & N/channel & NDIT & DIT[sec] & camera & Strehl\\
\hline
1 & 073.B-0084(A) & 2004-06-12 & H & Wollaston & 0, 45 & 30 & 1 & 30 & S13 & 0.171\\
2 & 083.B-0031(A) & 2009-05-18 & Ks & Wollaston & 0, 45 & 143 & 4 & 10 & S13 & 0.272\\
3 & 179.B-0261(A) & 2007-04-01 & Ks & Wollaston (rotated) & 0, 45 & 15 & 2 & 15 & S13 & 0.126\\
4 & 179.B-0261(A) & 2007-04-03 & Ks & Wollaston (rotated) & 0, 45 & 70 & 2 & 15 & S13 & 0.217\\
5 & 179.B-0261(A) & 2007-04-04 & Ks & Wollaston (rotated) & 0, 45 & 23 & 2 & 15 & S13 & 0.173\\
6 & 179.B-0261(A) & 2007-04-05 & Ks & Wollaston (rotated) & 0, 45 & 70 & 2 & 15 & S13 & 0.103\\
7 & 179.B-0261(A) & 2007-04-06 & Ks & Wollaston (rotated) & 0, 45 & 51 & 2 & 15 & S13 & 0.138\\
8 & 179.B-0261(D) & 2007-07-18 & Ks & Wollaston (rotated) & 0, 45 & 130 & 2 & 15 & S13 & 0.188\\
9 & 179.B-0261(D) & 2007-07-19 & Ks & Wollaston (rotated) & 0, 45 & 70 & 2 & 15 & S13 & 0.075\\
10 & 179.B-0261(D) & 2007-07-20 & Ks & Wollaston (rotated) & 0, 45 & 70 & 2 & 15 & S13 & 0.084\\
11 & 179.B-0261(D) & 2007-07-20 & Ks & Wollaston (rotated) & 0, 45 & 70 & 2 & 15 & S13 & 0.137\\
12 & 179.B-0261(D) & 2007-07-21 & Ks & Wollaston (rotated) & 0, 45 & 51 & 2 & 15 & S13 & 0.137\\
13 & 179.B-0261(D) & 2007-07-23 & Ks & Wollaston (rotated) & 0, 45 & 124 & 2 & 15 & S13 & 0.079\\
14 & 179.B-0261(D) & 2007-07-23 & Ks & Wollaston (rotated) & 0, 45 & 41 & 2 & 15 & S13 & 0.023\\
15 & 179.B-0261(D) & 2007-07-24 & Ks & Wollaston (rotated) & 0, 45 & 30 & 2 & 15 & S13 & 0.081\\
\hline
\end{tabular}
\end{table*}
\subsubsection{IRS~10}
IRS~10W shows a flux variability in all three bands, but there is no clear periodic behavior (see Fig.\ref{FigIRS10Wfluxvar},
only values not affected by saturation are shown). The Ks-
and H-band fluxes vary by about $\sim$30\%, while the L-band flux only shows about 10\% variability. We do not consider this a reliable
detection of intrinsic variability, because no systematic trends are visible at the observable timescale. By comparison, IRS~10E shows a
strong and clear periodic variability with the flux increasing by a factor of 3 and a period of about two
years (see Fig.\ref{FigIRS10Efluxvar}). The variability is correlated in the H-, Ks and L-band. This source has been classified as a
late-type Mira variable by \cite{tamura1996}, while \cite{ott1999} label it as a long-period variable. Our measurements agree with
these classifications.
\subsubsection{IRS~5}
IRS~5 also shows a flux increase in the Ks-band between 2002 and 2008 (see Fig.\ref{FigIRS5fluxvar}, one value affected by
saturation was taken out), although the
increase is not as strong and clear as that found for IRS~21. The H-band data is much less clear for this source,
and while it appears to be variable at that wavelength as well, the variability seems quite erratic. We therefore cannot
conclude that this is an intrinsically variable source on the observed timescale.
\section{Conclusions}
\label{SectDiscussion}
We draw the following conclusions:\\
\begin{enumerate}
\item Like several previous polarimetric surveys of the central parsec at much lower resolution, we find that the
polarization follows the Galactic plane in general. This confirms that the polarization can indeed be attributed to
aligned dust grains between the Galactic Center and the observer (foreground polarization). We find systematic
and individual deviations from this general pattern, however. Over our FOV we
find that the polarization degrees and angles change toward the eastern region around IRS~1. This behavior occurs in
the H- and the Ks-band data, with stronger relative deviations in the Ks-band.  That can be expected if this is indeed a local
effect produced by dust grains in the Northern Arm, because typical temperatures and grain sizes in this region lead to a much
more significant impact of radiation from this grain population at longer wavelengths. This is supported by a systematic change of
$\frac{p_H}{p_{Ks}}$ over the FOV, with values of 2.0$\pm$0.3 in the center and 1.6$\pm$0.3 in the Northern Arm region.
The higher of these values leads to a polarization maximum at $\sim 0.7 \mu$m, which matches previous findings of $\lambda_{max}
\sim 0.8 \mu$m by \cite{bailey1984}, based on JHK observations of a more extended region in the GC. This value represents dust parameters,
specifically typical grain sizes, and it can be expected that local contributions by distinct dust masses lead to a different
value than the average along the LOS. Unfortunately, we cannot derive any more constraints for the local dust based only on
these two bands, other than that the dust grains responsible for the local effects must be elongated and aligned
\citep[see][]{aitken1998}, and that their temperature and size must be sufficient for their influence in the H/Ks-band.  
\item We find a correlation between the spatially variable extinction toward the central parsec and the polarization efficiency,
$\frac{p_{\lambda}}{A_{\lambda}}$. This relation appears to be different for the two groups of sources we find in our FOV (sources
only affected by LOS polarization respectively sources where local contributions are significant): for the foreground-polarized
sources, we find a power law relation with a similar power law index for the H- and the Ks-band ($\beta_H = -0.6 \pm 0.6$,
$\beta_{Ks} = -0.4 \pm 0.4$ ). This is compatible with the measurements resp. the model
by \cite{jones1989}, who assume a combination of a constant and a random component of the galactic magnetic field along the
LOS to align the dust grains. For the population of higher polarized sources, we find a higher polarization efficiency in general,
while the power law index for both bands resembles that of the sources with lower polarization (with $\beta_H = -0.4 \pm 0.7$ and
$\beta_{Ks} = -0.5 \pm 1.5$, although it has to be cautioned that these fits are poor because of the small number of sources).
The higher polarization efficiency points to the influence of local dust grains, which only contribute
a small amount to the total extinction. But because these grains are aligned, they add a significant component to the total
polarization, because this alignment is not (partially) compensated by averaging over other dust populations with different
alignment as is the case for every individual dust configuration along the LOS. These findings further support our previous
conclusion that the influence of local dust is quite significant not only in the MIR, but also in the H- and Ks-band.
\item Intrinsic polarization takes place in several sources in the central parsec, not only at longer wavelengths as shown
first by \cite{knacke1977}, but also in the H- and Ks-band \citep[e.g.][]{ott1999}. With the M\"uller calculus we can
isolate the intrinsic component for point sources and maps of extended features. The resulting total intrinsic
polarization angles for IRS~21 and IRS~1W agree very well with what can be expected from source morphology and relative motion of
gas/dust in the Northern Arm and the sources themselves. We find very similar intrinsic polarization degrees for both sources,
with (7.8 $\pm$ 0.5) \% at (-75 $\pm$ 5)$^{\circ}$ for IRS~1W and (6.1 $\pm$ 0.5) \% at (5 $\pm$ 5) for IRS~21 (both Ks-band).
Contrary to the wavelength dependency seen in the foreground polarization, we find a slightly lower H-band polarization degree
for IRS~1W compared to the Ks-band: (6.9 $\pm$ 0.5) \% at (-73 $\pm$ 5)$^{\circ}$. That the central source contributes
a larger amount of flux in the H-band compared to the extended component, while the intrinsic polarization mostly stems from
the bow-shock, explains this discrepancy.\\    
The spatially resolved intrinsic polarization patterns of both bright extended sources in our FOV do not show a
centrosymmetric shape as would be expected for Mie scattering on spherical grains. Instead, the pattern found for IRS~1W
suggests that scattering on aligned elongated grains must play a significant role, possibly combined with thermal dust emission
(in the Ks-band), because the grains in the bow-shocks may reach temperatures on the order of 1000 K \cite[see][]{moultaka2004}.
Both effects would produce polarization parallel to the long axis of the grains and thus perpendicular to their angular momentum.
It therefore appears that the angular momentum of the grains in the bow-shock surrounding IRS~1W is aligned parallel to the
relative motion of the source through the Northern Arm material. Taking the magnetic fields in the Northern Arm into account,
this is most likely caused by the field lines being warped around the flanks of the bow-shock. This in turn compresses the field
lines and thus increases the local field strength, which is then sufficient for fast magnetic alignment of the grains.\\
For IRS~21 we find a different polarization pattern. It appears that the polarization vectors are mostly parallel to the relative
motion of the source through the surrounding material. Combined with the local field orientation, this again leads to the same dominant
mechanism of scattering/emission as for IRS~1W, if we assume that the field lines are compressed in front of the shock that
is moving perpendicular to them.
\item After deconvolution, IRS21 clearly does not show a circular shape in the Ks-band. Instead, its shape is consistent
with that of a poorly resolved bow-shock with its apex north of the central source. This is consistent with the L-band
shape presented by \cite{tanner2004} and the expectations from the proper motions of the source in relation to the streaming
motion of the Northern Arm. We find a standoff distance of $\sim$400 AU, this agrees with the upper limit given by
\cite{tanner2002}. It is difficult to constrain the nature of the central source by this parameter alone, and most stellar
types with strong winds, such as Wolf-Rayet stars, would be likely candidates \citep[see][]{tanner2002}. 
\item We find a flux variability for several bow-shock sources in the central parsec: IRS~21 shows a flux increase
of $\sim$50\% over 6 years in the H- and Ks-band (less, but still significant in the Lp-band). Especially the H-band measurement
indicates that this variability must be intrinsic to the central source and not only to the envelope. The candidates suggested
for the central source by \cite{tanner2002} could all show this variability, especially a source that is currently in a mass-losing
phase.\\
While we also find a possible flux variability for IRS~5, IRS~1W, and IRS~10W in all three bands, no clear periodicity or trend
is apparent for these sources at our time resolution. We therefore cannot determine if these sources show a true intrinsic variability
or not. We confirm the known periodicity of IRS~10E ($\sim$2 yrs) in the Ks-band, finding the same variability in the H- and Lp-band
as well.
\end{enumerate}
Several effects appear to contribute to the observed polarization in the central parsec of the GC, in addition to the foreground
polarization. The findings in our relatively narrow field-of-view alone indicate possible additional areas of interest: do the
large-scale trends we find continue in other sections of the Northern Arm (and the other elements of the minispiral)? Do the other bow-shock sources, such as IRS~10W, 5, and 8 show similar polarization patterns, and what does that tell us about the local processes in 
the bow-shocks? Are other intrinsically polarized sources present in the field, and what type of objects would show this behavior?
More observations with wider FOVs and in different bands (especially L-band, to directly examine extended dust emission e.g. in
the Northern Arm) would be desirable, and the recently achieved direct polarimetric calibration of NACO offers the possibility to map
more extended areas of the GC in polarization at high resolution.
\begin{acknowledgements}
We are grateful to all members of the NAOS/CONICA and the ESO PARANAL team, especially J. Girard. R. M. Buchholz acknowledges
support by Renia GmbH. R. Sch\"odel acknowledges support by the Ram\'on y Cajal programme, by
grants AYA2010-17631 and AYA2009-13036 of the Spanish Ministry of Science and Innovation, and by grant P08-TIC-4075 of the Junta
de Andaluc\'ia. The authors also thank the anonymous referee for the helpful comments.
\end{acknowledgements}

\begin{appendix}
\section{Error estimation}
\label{SectionErrors}
\begin{figure}[!b]
\centering
\includegraphics[width=\textwidth,angle=0, scale=0.50]{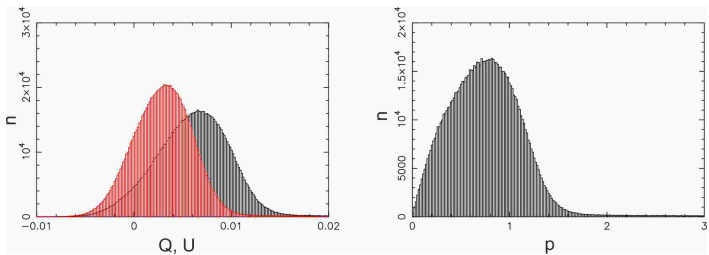}
\caption{\small Stokes parameters Q and U (left frame) and resulting polarization degree (right frame), measured
for each pixel in a wiregrid flat-field.}
\label{FigFlatfieldWG}
\end{figure}
\begin{figure*}[!t]
\centering
\includegraphics[width=\textwidth,angle=-90, scale=0.65]{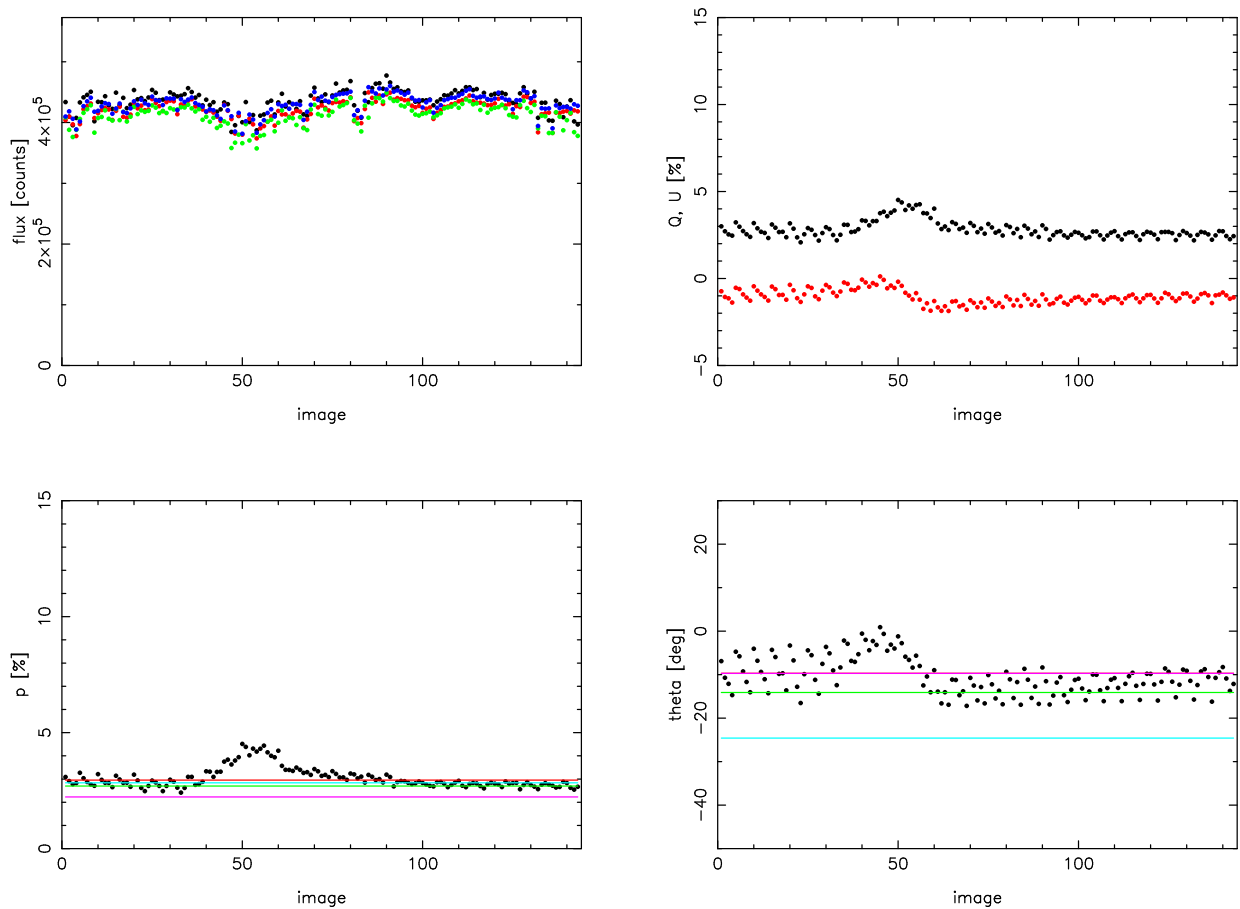}
\caption{\small Fluxes and polarization parameters of IRS~16C plotted against time of observation. Upper left:
flux measured in four channels. Upper right: Q (black) and U (red). Lower left: polarization degree. Lower right:
polarization angle. The plotted lines represent the polarization parameters obtained on the complete mosaic by aperture photometry 
(red), PSF fitting photometry using the IRS~16 stars (light blue) resp. the IRS~1 stars (magenta), resp. the final value after
deconvolution assisted PSF fitting photometry (green).}
\label{FigIRS16Ccomp}
\end{figure*}

\begin{figure*}[!t]
\centering
\includegraphics[width=\textwidth,angle=-90, scale=0.65]{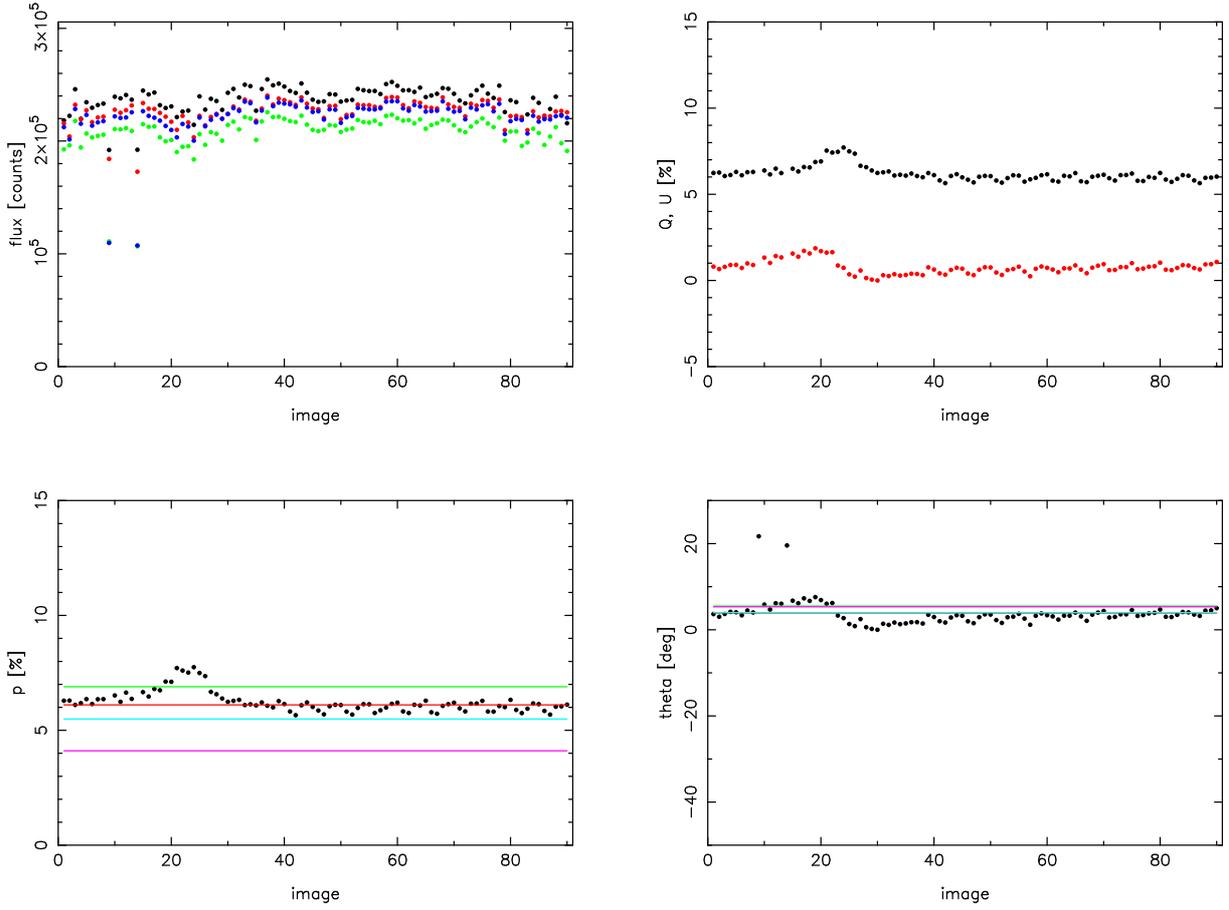}
\caption{\small Fluxes and polarization parameters of IRS~1C plotted against time of observation. Upper left:
flux measured in four channels. Upper right: Q (black) and U (red). Lower left: polarization degree. Lower right:
polarization angle. The plotted lines represent the polarization parameters obtained on the complete mosaic by aperture photometry 
(red), PSF fitting photometry using the IRS~16 stars (light blue) resp. the IRS~1 stars (magenta), resp. the final value after
deconvolution assisted PSF fitting photometry (green).}
\label{FigIRS1Ccomp}
\end{figure*}

 
In general, several effects can contribute to the total flux uncertainty: the basic uncertainty in the counts measured for each
pixel, the insufficiently determined PSF (resp. the fact that a PSF determined from several bright sources does not fit every
source perfectly) and possible effects of insufficient calibration (e.g. errors in the sky or the flat-field).\\
When using StarFinder, the formal flux error given by the algorithm represents the first of these three contributions. The
algorithm assumes an independence of all the pixels resp. their measured counts, however. This is not the case anymore for a
deconvolved image. \cite{schoedel2010a} circumvented this problem by introducing a modification factor for these formal errors
(determined by simulations). Owing to the different instrument setup used here, we could not simply adopt the same factor.\\
\begin{figure}[!t]
\centering
\includegraphics[width=\textwidth,angle=-90, scale=0.35]{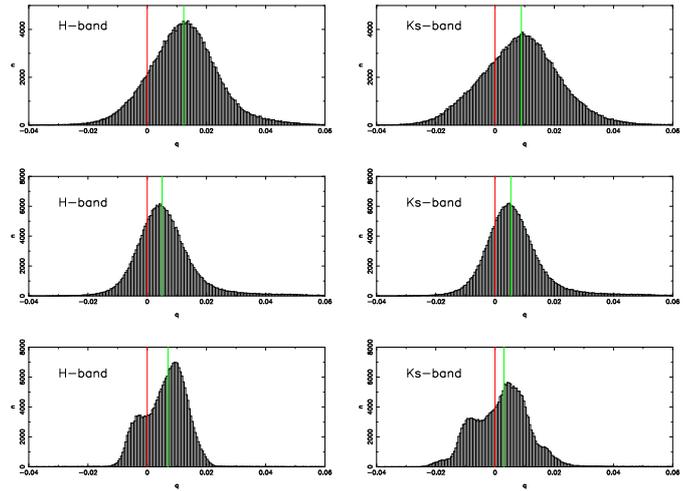}
\caption{\small Stokes parameter Q measured at each pixel of a flat-field taken with (upper frames) and without (middle
frames) the Wollaston prism. The lower frames show Q after dividing the Wollaston flat-field by the flat-field taken
without the Wollaston. Left frames: H-band, right frames: Ks-band.}
\label{FigFlatfieldWoll}
\end{figure}
The second contribution stems from a variation of the PSF over the FOV.
Generally, the shape of the PSF changes with the relative
position of the source to the guide star (anisoplanasy), with longer projected distances leading to larger distortions. The
extent of this behavior is highly dependent on the data quality resp. the seeing. The correlation factor provided by StarFinder
describes this effect, because it measures how well a source matches the PSF,
but it is not clear how to translate this into a flux
uncertainty. We therefore estimated this effect in another way: we extracted three different PSFs from the western, central, and
eastern region of the FOV and used them for PSF fitting with StarFinder. As it turned out, this does not only produce different
fluxes in all four channels, but can also lead to a shift in polarization angle and/or degree (depending on the source). A similar
effect can be seen when comparing fluxes and polarization parameters measured in different sub-frames of the deconvolved image.\\
We compared these measurements to aperture photometry applied to two bright sources (IRS~16C, IRS~1C). The results
are shown in Figs.\ref{FigIRS16Ccomp},\ref{FigIRS1Ccomp}. In general, we find that the values determined
by aperture photometry are matched better by the PSF-fitting values the closer the PSF reference sources are to the source in
question. The deviations can reach values of up to 1.5\% resp. 10$^{\circ}$ even for these bright sources.\\
The reason for this behavior is revealed by a closer look at the individual PSFs: these show a polarization themselves, and that
pattern changes over the FOV (see Fig.\ref{FigPSFpol}, the polarization vectors are almost perpendicular in several locations,
even in areas with significant flux). This leads us to conclude that the fluxes determined with a ''nearby'' PSF are more accurate
than those determined with a ''distant'' PSF, and we therefore apply a weighted average (based on the distance of the individual source
to the average position of the PSF stars) to the photometry on the deconvolved sub-images. The results provide a better match to
the aperture photometry (though it has to be cautioned that the latter is problematic in crowded regions and where large-scale
extended emission is present, which is why this method is not applied to the image as a whole). The standard deviation of the
fluxes over the sub-frames provides an estimate for the contribution of this effect to the total uncertainty.\\ 
Using these error contributions to estimate the final, total uncertainty is problematic: the sub-frames are all taken from the same
mosaic and are thus not independent measurements, the formal errors given by StarFinder are unreliable and generally far too small
for brighter sources, and possible calibration problems (sky, flat-field) are not covered at all.\\
We therefore used a different approach to determine the total flux errors: we applied the same photometric method that we used on
the full mosaic to mosaics created from eight (Ks-band) resp. six (H-band) sub-datasets, selected based on the dither position
along the East-West axis. This led to
sub-sets of about 15 (Ks-band) resp. 5 (H-band) images per dither position, and the resulting mosaics have the same depth
resp. signal-to-noise ratio
everywhere. Because the original images can be regarded as independent measurements, so can the resulting mosaics, with the
additional advantage that fainter stars still have a sufficient signal-to-noise ratio (as compared to individual images). The
downside of this approach is that all sources in the eastern- and westernmost $\sim$2'' are contained in only one of these
sub-mosaics. We then adopted the standard deviation of the fluxes measured in these sub-mosaics as the photometric error of each
source.\\
In several cases (mostly in the H-band) this method yielded extremely small errors for weak sources
(see Fig.\ref{FigPhoterrsHK}). This can happen by coincidence if a source is only contained in a small number of sub-mosaics.
We consider these errors (and the photometry of these sources in general) unreliable, and the respective source will have a
tendency not to pass the reliability criterion for the polarimetry.\\
The aperture photometry applied to individual frames shown e.g. in Fig.\ref{FigIRS16Ccomp} reveals two other effects: the measured
flux varies over the frames, and in addition to the instrumental polarization (most prominently the ''bump'' in p), small variations
in Q and U (and subsequently p and $\theta$) can be observed. These variations correspond to the dither pattern used in the
observation, and they probe the different areas of the Wollaston prism resp. the detector. These deviations are much smaller than
the photometric errors, however, so we assume that we can safely neglect resp. average over them.\\
\cite{witzel2010} determined the system-intrinsic systematic uncertainty of polarimetric observations with NACO as 1\% in
polarization degree and 5$^{\circ}$ in polarization angle. For a source with 5\% polarization (a typical value for GC sources in
the Ks-band), this already corresponds to a relative error of 20\%. We simulated the required accuracy of the photometry so that
the final polarimetric errors stay below three times (resp. six times for the H-band) the systematic uncertainty.
We tolerate higher
uncertainties in the H-band because the higher polarization there leads to a lower relative error for the same absolute uncertainty
of the polarization degree. This led to error thresholds of 3\% in the Ks-band and 6\% in the H-band. Most of the sources brighter 
than 16 resp. 18 mag (Ks/H) have errors below these thresholds (see Fig.\ref{FigPhoterrsHK}). This leads to the final error
of the polarization parameters being dominated by the statistical photometric errors.
\section{LR deconvolution: useful for extended structures?}
\label{SectSimLR}
\begin{figure*}[!t]
\centering
\includegraphics[width=\textwidth,angle=-90, scale=0.65]{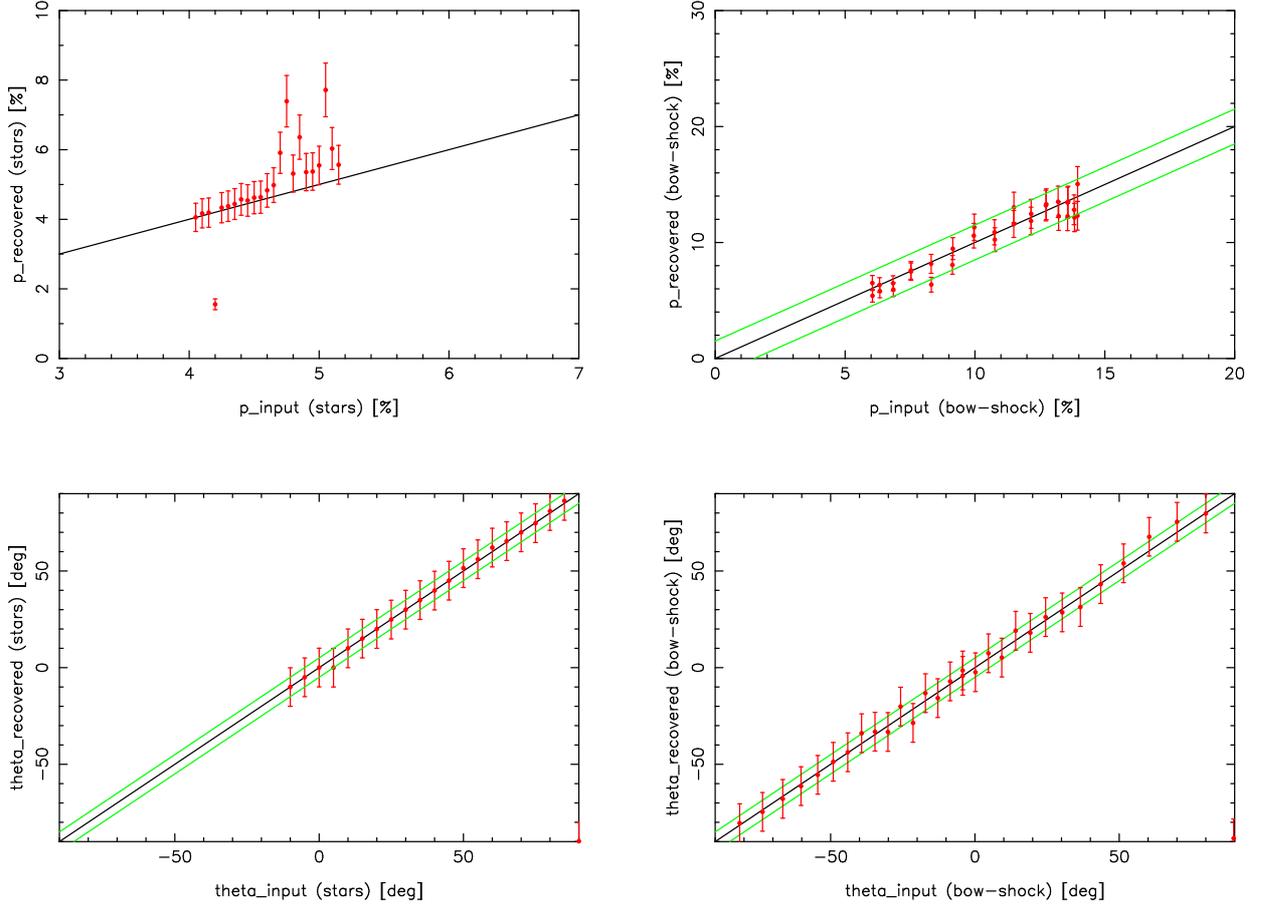}
\caption{\small Recovered polarization parameters for point sources (left frames) and extended feature (right
frames). The green lines indicate a deviation of $\pm$1.5\% resp. 5$^{\circ}$ from a perfect recovery, indicated by
the black line.}
\label{FigSimLRdec}
\end{figure*}
It is known that Lucy-Richardson deconvolution has a tendency to ''suck up'' surrounding flux into a central
point source \citep[see e.g.][]{schoedel2010a}. This impedes or severely hinders the detection of faint sources
close to bright ones, introduces astrometric and photometric uncertainties and makes a reliable background
estimation impossible. But unlike linear deconvolution, this method allows flux measurements on extended structures
via simple aperture photometry on the deconvolved images. Can it be used to measure the polarization of the
extended sources in the GC?\\
To investigate this, we produced images of an artificial bow-shock-like structure next to a point source.
Several other point sources with different fluxes were added to simulate a crowded field as is found in the
GC. For each filter (0$^{\circ}$, 45$^{\circ}$, 90$^{\circ}$ and 135$^{\circ}$), we modified all fluxes by a factor calculated for generic
polarization parameters of P = 4-5\% and $\delta$ = -10$^{\circ}$-100$^{\circ}$:
\begin{equation}
f_{pol,rel} = 1 + P \times cos(\theta-\delta),
\end{equation}
This was applied to the extended feature as well, with a constant polarization degree of 10\% and the polarization angles
parallel to the bow-shock structure. We then added an additional foreground polarization of 4\% at -15$^{\circ}$ to
the extended feature. The resulting map was convolved with a PSF extracted from the real images. We also added a
weakly position-dependent background in the shape of a very flat two-dimensional Gaussian similar in level to that
found in the actual observations.\\
\begin{figure}[!b]
\centering
\includegraphics[width=\textwidth,angle=-90, scale=0.45]{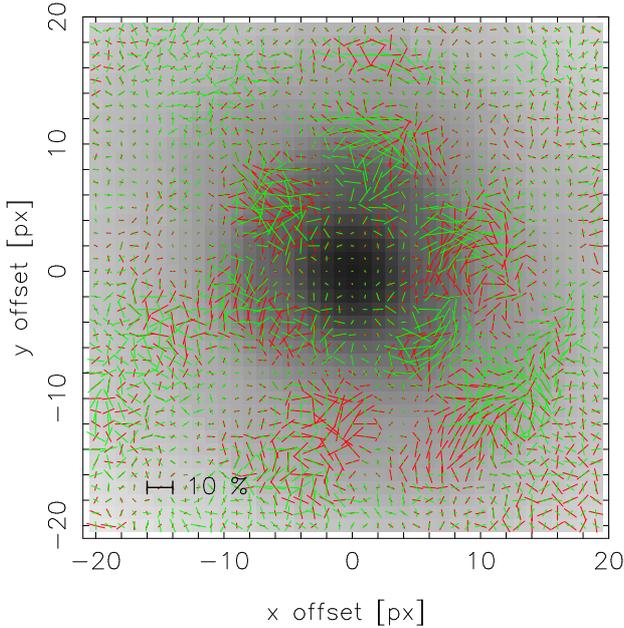}
\caption{\small Polarization pattern of the PSF used for the first step of the deconvolution-assisted photometry. Red: PSF determined 
from IRS~16 sources. Green: PSF determined from IRS~1 sources.}
\label{FigPSFpol}
\end{figure}
We applied the Lucy-Richardson deconvolution algorithm to these artificial images, followed by aperture photometry on the point
sources and the extended structure, mapping the latter with two pixel apertures along its length. The structure itself was
successfully recovered, although slightly widened. Fig.\ref{FigSimLRdec} shows a comparison of the
recovered polarization parameters in comparison to the input values. It appears that the only significant large
deviation occurs for the source in the center of the extended feature: the recovered polarization degree is 2.6
percentage points lower than the input value, while the polarization angle is 5$^{\circ}$ less. This appears to be an
effect of the algorithm ''sucking'' flux of the extended feature into the central source, thereby changing its
polarization because of the significantly different polarization of the shell. The polarization degrees of several of
the very faint point sources were also recovered poorly, while at the same time their recovered polarization angles
agreed much better with the input values.\\
The values recovered for the bow-shock-feature itself
show only small deviations compared to the input values, with a maximum difference in polarization degrees of
1.5-2\% (smaller in most cases) and 5-7$^{\circ}$ in polarization angle. The deviations are even smaller for the
brighter stars, with less than 0.5\% and less than 1$^{\circ}$ in most cases. The fainter stars show only slightly
larger deviations. This approaches the systematic uncertainties inherent to the instrument. We can therefore consider
the Lucy-Richardson algorithm to be sufficiently accurate for our purposes of detecting extended features. The small
variances found for the artificial ''bow-shock'' also indicate the level of accuracy that can be expected for
the polarization parameters determined in \S \ref{SectExtended}.\\
Fig.\ref{FigPSFpol} shows the polarization pattern of two PSFs determined from different bright sources in the FOV (IRS~16
and IRS~1 point sources). There are significant spatial variations, so these variations must occur (to some extent) in all
sources in the non-deconvolved images. Because these sources are expected to be essentially point-like before their light
enters the atmosphere, polarization patterns like this must be introduced by instrumental or atmospheric effects. This
is another reason why deconvolution is an important step to recover polarization patterns of extended sources
successfully: the intrinsic patterns would otherwise be masked by the strong spatial variations, with local values for the
polarization degree of up to 15\% in the PSF pattern (although these values are much lower in the central region with the
highest flux, on the order of 1-2\%). These effects are removed by the deconvolution process.\\
\end{appendix}
\end{document}